\documentclass{article}
\usepackage{arxiv}
\usepackage[utf8]{inputenc}
\usepackage{verbatim}
\usepackage{lscape} 
\usepackage{rotating}
\usepackage{multirow}
\usepackage{color,soul}
\usepackage{url}
\usepackage{enumitem}

\usepackage{comment}
\usepackage{threeparttable}

\begin{document}

\title{Demystifying COVID-19 digital contact tracing:\\
A survey on frameworks and mobile apps}

\author{
 Tania Martin, Georgios Karopoulos, José L. Hernández-Ramos, Georgios Kambourakis, and Igor Nai Fovino \\
  European Commission\\
  Joint Research Centre\\
  Ispra 21027, Italy \\
  \texttt{\{tania.martin, georgios.karopoulos, jose-luis.hernandez-ramos,}\\ \texttt{georgios.kampourakis, igor.nai-fovino\}@ec.europa.eu} \\
}

\maketitle

\begin{abstract}

The coronavirus pandemic is a new reality and it severely affects the modus vivendi of the international community. In this context, governments are rushing to devise or embrace novel surveillance mechanisms and monitoring systems to fight the outbreak. The development of digital tracing apps, which among others are aimed at automatising and globalising the prompt alerting of individuals at risk in a privacy-preserving manner is a prominent example of this ongoing effort. Very promptly, a number of digital contact tracing architectures has been sprouted, followed by relevant app implementations adopted by governments worldwide. Bluetooth, and specifically its Low Energy (BLE) power-conserving variant has emerged as the most promising short-range wireless network technology to implement the contact tracing service.
This work offers the first to our knowledge, full-fledged review of the most concrete contact tracing architectures proposed so far in a global scale. This endeavour does not only embrace the diverse types of architectures and systems, namely centralised, decentralised, or hybrid, but it equally addresses the client side, i.e., the apps that have been already deployed in Europe by each country. There is also a full-spectrum adversary model section, which does not only amalgamate the previous work in the topic, but also brings new insights and angles to contemplate upon.

\end{abstract}

\section{Introduction}


The World Health Organization (WHO) on March 11, 2020 declared COVID-19 a pandemic\footnote{https://www.who.int/dg/speeches/detail/who-director-general-s-opening-remarks-at-the-media-briefing-on-covid-19---11-march-2020}, whose effects will probably determine the evolution of our society for many years to come. The direction of this evolution will greatly depend on the capacity of our society to swiftly and jointly converge toward the best mitigation solutions. Until a vaccine will be available or unless the pandemic will spontaneously disappear, the best weapons in the hands of countries will be prevention and fast diagnosis of infected people. Indeed, in the global race against the spread of the COVID-19, countries, public and private organisations, the academia, and others have quickly joined the forces to orchestrate appropriate countermeasures.

In this context, the development of contact tracing approaches is currently considered as one of the main weapons to confront the spread of the COVID-19 worldwide. Indeed, contact tracing is considered by the WHO as a key component of the infection monitoring by including contact identification, listing and follow-up aspects~\cite{WHO2017Contact}. So far, contact tracing has been mainly based on manual procedures, in which infected people are interviewed in an attempt to trace their contacts. Then, the health authority reaches each contact to check if they present any symptoms and advises them accordingly, e.g., get tested and/or self-quarantine. This approach is time consuming, resource demanding, and prone to errors, since people might not remember all their contacts or, even if they do, they might not know them in person or how to contact them. 

To cope with these issues, digital contact tracing has emerged with different initiatives, which are currently driven by organisations and governments worldwide. The main purpose is to efficiently detect people who have been in close contact with infected individuals, so they can be promptly and properly advised on the next steps to follow. This way, potentially infected individuals can be easily detected and self-isolated even before showing symptoms. Therefore, the infection chain is interrupted as early as possible. 

During the last few months, numerous contact tracing frameworks and smartphone  applications (apps) have emerged. These frameworks comprise the backend infrastructure and the protocols used to communicate among subsystems, whereas the apps are installed on peoples' smartphones and interact with the backend infrastructure. However, the development of such frameworks and apps poses security and data protection issues, in addition to interoperability concerns. Indeed, at the European Union level, these aspects are highlighted by recent legal instruments, such as the EU Recommendation 2020/518~\cite{covid_recommendation} and the Commission Communication 2020/C~124~I/01~\cite{covid_communication}.
Furthermore, in the current COVID-19 era, an increased number of fraudulent activities related to the pandemic has been observed~\cite{CovidCyberIncidents}.
Although there is rich research work on protecting data in the healthcare sector~\cite{BlockchainEhrs}, the emergence of contact tracing apps processing sensitive data in the end-user device requires a different approach to identify potential vulnerabilities~\cite{kouliaridis2020dissecting}.


\emph{Our contribution:} Taking into account the current landscape of digital contact tracing frameworks and apps, this work endeavours to provide a comprehensive overview of such efforts and to analyse the main security and data protection aspects around these initiatives. In particular, we scrutinise recent frameworks jointly developed by industry and academia, such as the Decentralised Privacy-Preserving Proximity Tracing (DP-3T)~\cite{DP3TDP3T} or the Pan-European Privacy-Preserving Proximity Tracing (PEPP-PT)~\cite{PEPP-PTPan-European}. Furthermore, we describe a full-fledged adversarial model, which brings new insights and angles to be considered for the development and evolution of ongoing contact tracing initiatives. Such a model is used to analyse the different frameworks around different security and data protection concerns. Finally, we provide an extensive overview of the main EU apps that are already deployed or currently in development in different countries. To the best of our knowledge, this is the first paper providing a sweeping overview of current contact tracing frameworks and mobile apps coping with the COVID-19 pandemic. We believe that our work could be used as a reference for researchers working in the definition of digital contact tracing approaches to restrain the spread of the COVID-19, as well as general contact tracing initiatives focused on security and data protection aspects. 

The remainder of this paper is organised as follows. Section~\ref{sec:frameworks} details on the main contact tracing frameworks developed so far. Section~\ref{sec:threat} offers an adversarial model that is well-suited to anatomise the security and privacy aspects of the various approaches. Furthermore, Section~\ref{sec:apps} explores the mobile apps already deployed or in development across the European continent. Finally, a conclusion is drawn in Section~\ref{sec:conclusions}.
\section{Digital contact tracing frameworks} \label{sec:frameworks}

As already mentioned, the definition of contact tracing approaches has attracted a significant interest recently. This section scrutinises the existing contact tracing frameworks, and analyses their chief operational aspects. The main centralised and decentralised frameworks used by most contact tracing apps are described in more detail, while a brief description of the rest is provided given that their operation is very similar to the former.

\subsection{Decentralized Privacy-Preserving Proximity Tracing (DP-3T)~\cite{DP3TDP3T}}\label{sec:dp3t}

The Decentralised Privacy-Preserving Proximity Tracing (DP-3T) represents a decentralised contact tracing approach, which is driven by several international experts from academia and research institutions. The DP-3T consortium was formed by several members of the Pan-European Privacy-Preserving Proximity Tracing (PEPP-PT) initiative (which is described in the next subsection) as a decentralised alternative, which is open source at a GitHub repository~\cite{DP3TDP3T}. According to the DP-3T team~\cite{Troncoso2020Decentralized}, the main objectives of the system are to enable a quick notification of contact people at risk, and to help epidemiologists to analyse the spread of the virus. Furthermore, the consortium has recently defined additional goals, including the communication with interested stakeholders to improve tracing systems, contributions about the effectiveness of tracing solutions, or collaboration for the development of related apps~\cite{DP3T2020Aims}. 

The DP-3T system is based on the broadcast of identifiers (IDs) through Bluetooth Low Energy (BLE) by the user's smartphone. Therefore, nearby users are enabled to receive and store such IDs. In case an infected person is detected, their smartphone is authorised to send their IDs to the backend, which in turn broadcasts the IDs to the users of the system. This way, each receiving user compares the received IDs against the list of stored IDs and, in case of an ID match, the app notifies the user that they have been in contact with an infected person. 

From an architectural perspective, the DP-3T system only requires a backend server and the users' smartphones, where the corresponding app is installed. Furthermore, the existence of a health authority is assumed. Then, the following two main processes are defined:

\begin{itemize}
    \item Generation and storage of ephemeral IDs (EphIDs).
    \item Proximity tracing.
\end{itemize}

\subsubsection{Generation and storage of ephemeral IDs (EphIDs)}

As already pointed out, the approach defines a core solution in which each smartphone broadcasts changing ephemeral IDs ($EphID$s), which are sent through BLE beacons (advertisements). These IDs are generated from a secret key $SK_t$, where $t$ represents the current day. Furthermore, the same key is refreshed every day by using a hash function $H$, in such a way that $SK_t = H(SK_{t-1})$. This is a hash chain scheme, meaning that if a key is compromised, then all the subsequent $SK$s are revealed, but not the $SK$s before it. Then, $SK_t$ is used to derive a set of $EphID$s by using a pseudo-random function $PRF$, say, HMAC-SHA-256, and a pseudorandom generator $PRG$, say, AES in counter mode:

\[EphID_1 || ... || EphID_n = PRG( PRF(SK_t, \textrm{``broadcast key''}))\]

\begin{figure}
\centering
\includegraphics[width=0.48\textwidth]{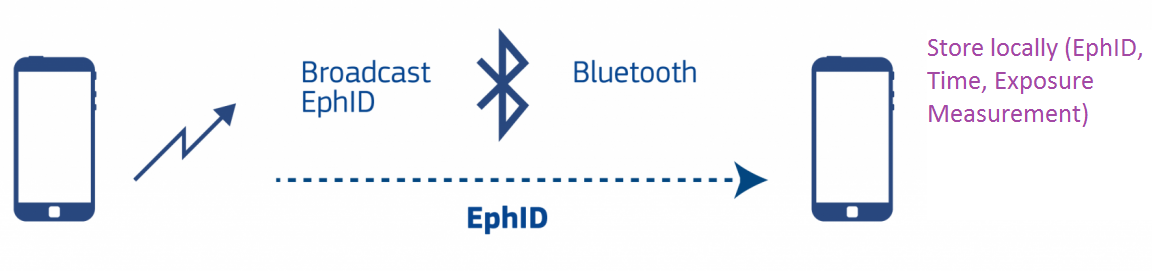}
\caption{DP-3T processing and storing of observed EphIDs ~\cite{Troncoso2020Decentralized}}
\label{fig:dp3t1}
\end{figure}

To avoid location tracking, each $EphID$ has a validity period of several minutes. $EphID$s are received by nearby users through BLE advertisements. Then, each $EphID$ is stored by these users together with an exposure measurement, e.g., signal attenuation, and the day when the beacon was received. This process is shown in Figure \ref{fig:dp3t1}. Furthermore, each user's app locally stores their own keys $SK_t$ that were generated during the past 14 days. 

\subsubsection{Proximity tracing}

\begin{figure*}
\centering
\includegraphics[width=0.9\textwidth]{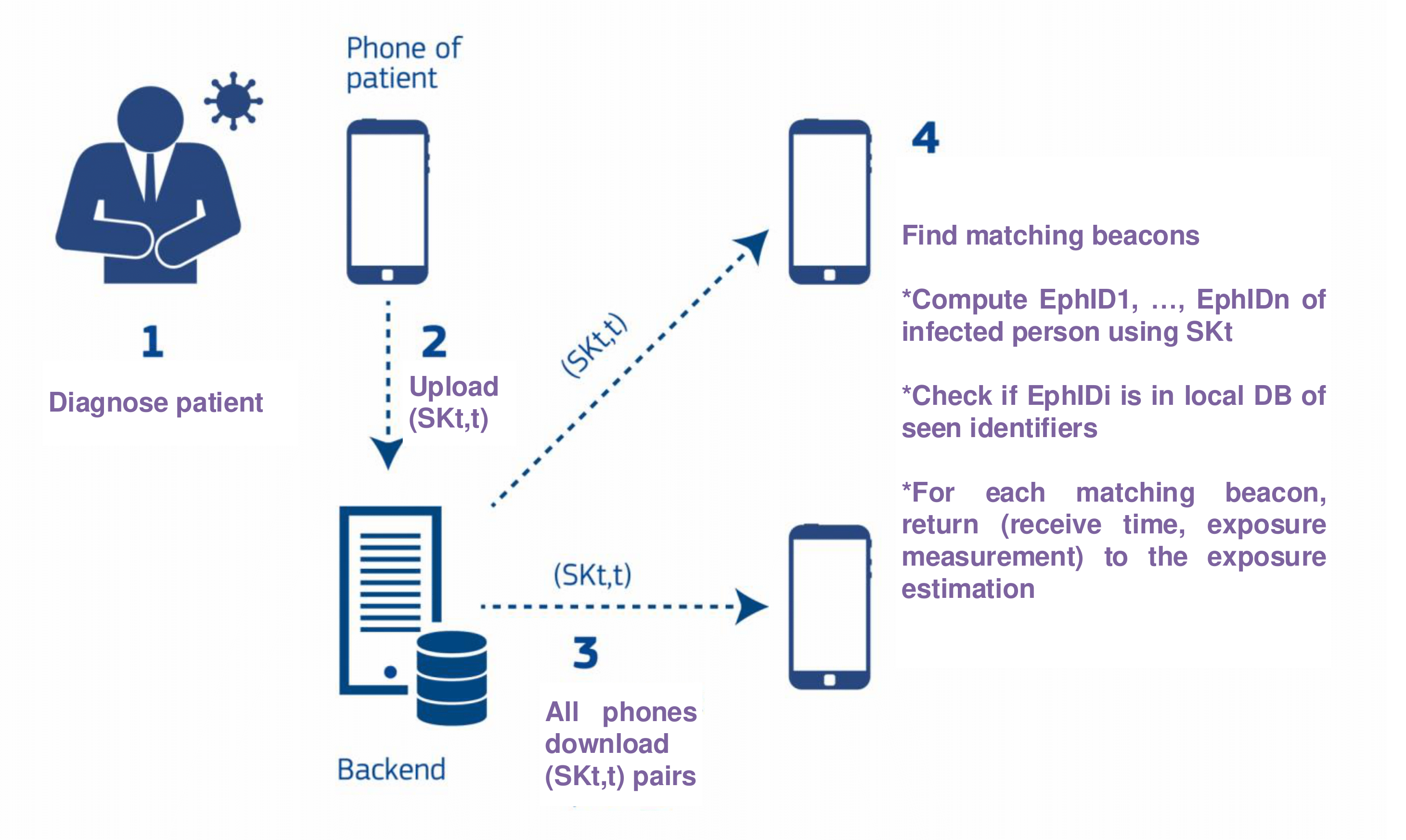}
\caption{DP-3T proximity tracing process  ~\cite{Troncoso2020Decentralized}}
\label{fig:dp3t2}
\end{figure*}

The process of proximity tracing illustrated in Figure \ref{fig:dp3t2} is triggered when a user is diagnosed as infected by the health authority. The latter authority is responsible for notifying test results \textcircled{1}, authorising users to upload information to the backend server, and calculating the time during a patient is contagious, also known as ``contagious window''. When a person is diagnosed as contagious and is authorised by the health authority, say, via an authorisation code, they upload the key $SK_t$ and the first day $t$ that they were considered to be contagious \textcircled{2}. This information can be encoded in the authorisation code. Therefore, the backend will receive a pair $(SK_t,t)$ of each infected individual. Then, the different $(SK_t,t)$ pairs are periodically downloaded by the registered users \textcircled{3}. It should be noted that the backend is only intended to broadcast this information, instead of processing any data. With this information, users are enabled to compute the list of $EphID$s associated to a given $(SK_t,t)$ pair. In case such an $EphID$ is included in their stored list, it means the user was in contact with an infected person. Then, for each matching beacon, the data on \textit{receive time} and \textit{exposure measurement} is sent to a exposure estimation component, which is intended to estimate the duration of the smartphone owner's exposure to infected users in the past.

The previous description pertains to the DP-3T design ``Low-cost decentralized proximity tracing''. However, the DP-3T consortium has also proposed an alternative approach called ``unlinkable decentralized proximity tracing''~\cite{Troncoso2020Decentralized}, which is intended to provide better privacy properties, but at the expense of stronger performance requirements on the smartphone side. In this case, when a user is diagnosed as infected, they can decide which IDs are shared to avoid the potential linking of $EphIDs$. For example, a user may choose not to share the IDs corresponding to a certain period, e.g., Sunday morning. The approach is based on the use of Cuckoo filters~\cite{Fan2014Cuckoo}, in which the IDs of an infected person are hashed and stored. 

Furthermore, a third alternative has been defined in the latest version of the DP-3T white paper \cite{Troncoso2020Decentralized}, which integrates the advantages of the aforementioned designs. Precisely, it is called ``Hybrid decentralized proximity tracing'' in which seeds are generated and used to create ephemeral IDs according to the first design, but these seeds are only uploaded in case they are relevant to exposure estimation for other users. This way, protection against linking ephemeral IDs is enhanced compared to the low-cost design, but tracking protection is weaker than for the unlinkable design. It should be noted that, according to DP-3T \cite{Troncoso2020Decentralized}, this design closely resembles the Google/Apple framework \cite{Apple2020ENExposureConfiguration} in which time windows are 1-day long, so one seed is used to generate the ephemeral IDs of that day. 

Moreover, the DP-3T consortium has proposed~\cite{Troncoso2020Decentralized} an enhancement that can be applied to diverse proximity tracing systems called ``EphID Spreading with Secret Sharing''. The main goal of this approach is to block an adversary from recording a proximity event, even in case the contact was during a very short period of time, or when the distance is actually long among people. Therefore, such an attacker could acquire a potentially large amount of $EphID$s that could be used to infer additional information about a certain user. To mitigate such an issue, the approach is based on splitting each $EphID$ into different shares, so that each share is transmitted using a certain BLE advertisement. On the downside, a potential receiver needs to get a minimum number of shares to be able to construct the corresponding $EphID$. 

\paragraph{Google/Apple Exposure Notification}
It should be noted that the solution proposed by Google and Apple, namely Exposure Notification~\cite{Apple2020Privacy-Preserving}, follows a decentralised approach, which was ``heavily inspired'' by DP-3T, according to Google~\cite{Etherington2020Apple}. Indeed, as already mentioned, the last version of DP-3T considers that this approach could be seen as a particular case of the ``Hybrid decentralized proximity tracing'' design. In Exposure Notification, the pseudorandom IDs that are broadcast over BLE, namely Rolling Proximity Identifiers (RPIs), are generated in a similar way as in DP-3T: a Temporary Exposure Key, which is changed every day, is used to derive the RPIs employing a hash function and the AES algorithm. The RPIs are renewed every time the BLE randomised address is changed, namely about every 15 min, to prevent linkability and wireless tracking. Whenever a user is diagnosed as positive to COVID-19, they share the latest Temporary Exposure Keys, e.g., covering the most recent 14 days, with a central server. The mechanism followed by an infected user to report the collected temporary IDs will be determined by their public health authority, say by using a one-time password, but this is not specified by the protocol. The server aggregates all the received Temporary Exposure Keys and the users of the Exposure Notification system periodically download this list of keys. If a user is never diagnosed as positive, their Temporary Exposure Keys do not leave the smartphone. The deployment of Exposure Notification has already started on May 25, 2020 with a large scale pilot in Switzerland~\cite{Barraud2020First}. Furthermore, other countries are also considering the use of this approach for their mobile apps, as described in Section~\ref{sec:apps}. For information about security considerations of the approach, the reader can refer to~\cite{Gvili2020Security}.

\subsection{Pan-European Privacy-Preserving Proximity Tracing (PEPP-PT)~\cite{PEPP-PTPan-European}}

The Pan-European Privacy-Preserving Proximity Tracing (PEPP-PT) is a digital proximity-tracing framework that uses BLE advertisements to discover and locally log to a user's smartphone other users that are in close proximity. According to its designers~\cite{PEPP-PT2020PEPP-PT-overview}, it notifies people at risk with a 90\% true positive and 10\% false negative rates. Initially, many different systems following either the centralised or decentralised approach were participating under this initiative, including DP-3T, whose partners eventually resigned from the PEPP-PT consortium. In the rest of this section, PEPP-PT refers to two very similar centralised systems, the PEPP-PT NTK~\cite{PEPP-PT2020PEPP-PT-german} and ROBERT~\cite{PRIVATICS2020ROBERT}. These systems employ a centralised reporting server to process contact logs and individually notify clients of potential contact with an infected patient. 
 

The PEPP-PT system comprises the following components:

\begin{itemize}

  \item A user mobile app for proximity tracing.
  \item A backend server for generating temporary IDs used with the app and processing the data received by the app.
  \item A push notification service\footnote{ROBERT does not include this component because it follows a pure pull approach where the app regularly checks the infection status of its user, in contrast to NTK where the backend requests from a subset of all users to check their status.} to trigger the app to pull notification from the backend.
\end{itemize}

\begin{figure*}
\centering
\includegraphics[width=15cm]{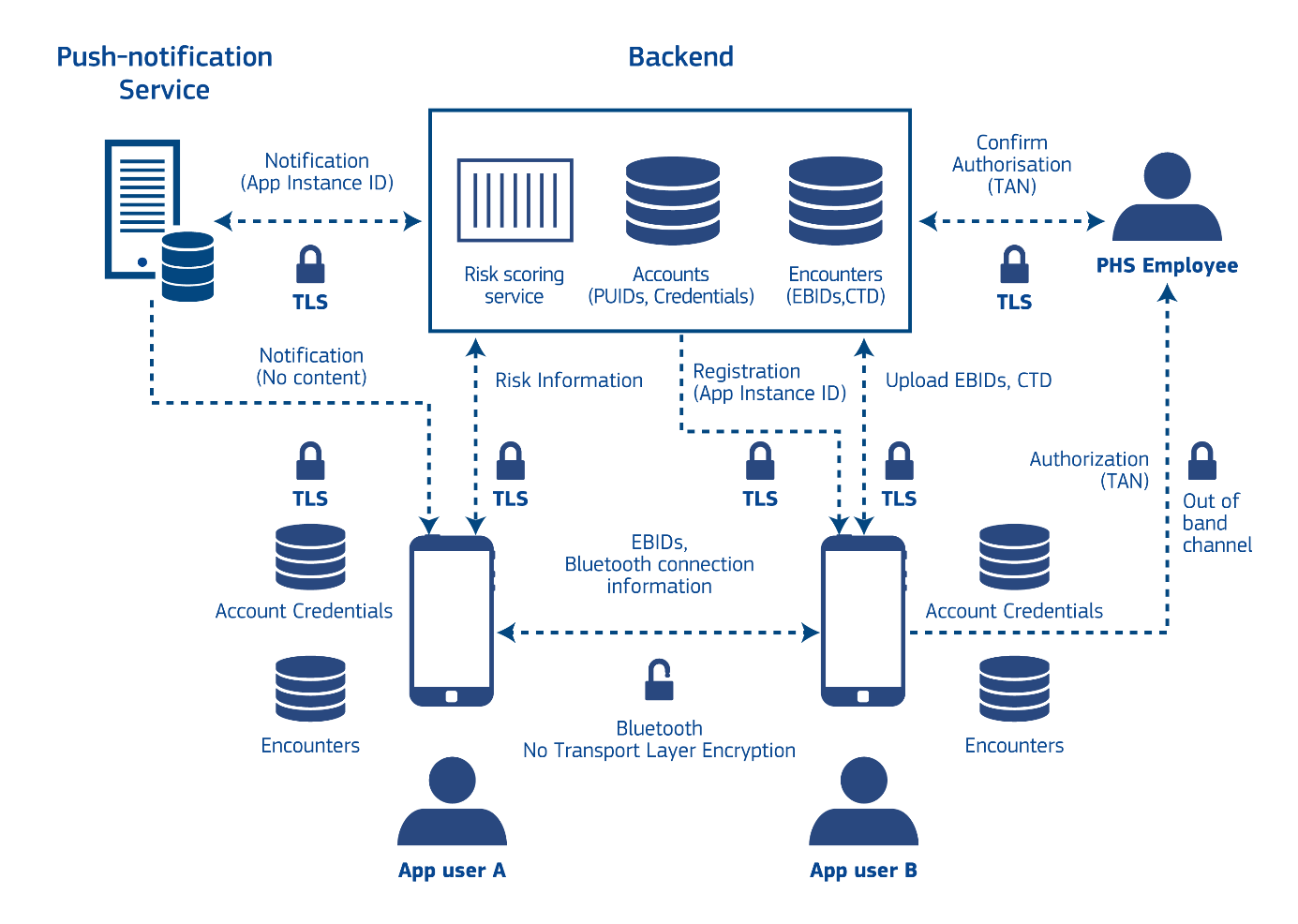}
\caption{High-level architecture of PEPP-PT NTK~\cite{PEPP-PT2020PEPP-PT-german}}
\label{fig:pepp-architecture}
\end{figure*}

An overview of the data stored in the different subsystems of PEPP-PT is presented in Table~\ref{table:pepp-data}. The interactions among the aforementioned components are depicted in Figure~\ref{fig:pepp-architecture}. These interactions are facilitated by the following protocols that will be analysed in the rest of this section:

\begin{itemize}

    \item User registration.
    \item Proximity-tracing of other smartphones.
    \item Sharing collected proximity data with the server.
    \item Federation with other backends.
    
\end{itemize}

\begin{table}
\caption{Data storage in PEPP-PT subsystems.}\label{table:pepp-data}
\centering
\renewcommand{\arraystretch}{1.5}
\begin{tabular}{|l|p{6.7cm}|}
\hline
\textbf{Subsystem} & \textbf{Data}\\
\hline
\hline
\multirow{7}{*}{Smartphone} & Set of current and future Ephemeral BLE IDs  ($EBID$s) to broadcast \\\cline{2-2}
& Proximity history of the last 21 days (containing the observed $EBID$s and timestamps)\\\cline{2-2}
& OAuth2~\cite{Hardt2012OAuth} client secret for access to backend services (long term)\\\cline{2-2}
& OAuth2 access token for access to backend services (short lived)\\
\hline
\multirow{6}{*}{Backend} & Persistent user ID  ($PUID$)  \\\cline{2-2}
& OAuth2 client credentials of an app\\\cline{2-2}
& OAuth2 temporary client access token (short term, 1 h)\\\cline{2-2}
& Medium term (days to weeks): backend keys ($BK_t$), $EBID$s, observed $EBID$ lists\\\cline{2-2}
& Push Notification Service ID (PID)\\
\hline
\multirow{2}{*}{Not stored} & Transaction Authentication Number (TAN): one-time password  for uploading the observed $EBID$ list to the backend\\
\hline
\end{tabular}
\end{table}

\subsubsection{User registration}

When a user installs a PEPP-PT-based app, the latter is always active in the background. During user registration, a pseudonymous user ID is generated by the server and sent to the app. Since attributes like email accounts and phone numbers are not used in PEPP-PT, a combination of a Proof-of-Work (PoW) and a Captcha is used in order to impede mass creation of user accounts. The PoW makes registrations quite expensive and prevents DoS attacks by unauthenticated requests, while Captcha requires human interaction. The registration steps are the following:

\begin{enumerate}

    \item The user requests to register to the backend.
    \item A PoW and a Captcha challenges are sent to the app.
    \item The app computes the solution to the PoW challenge and the user solves the Captcha.
    \item The two results are sent to the backend and verified.
    \item The app receives OAuth2 client credentials, i.e., random client ID and client secret.
    \item The backend stores the app's OAuth2 client credentials, a unique 128-bit random pseudonymous persistent user ID ($PUID$) and a push notification ID (PID).

\end{enumerate}

After registration, when the app needs to communicate with the backend, it uses its OAuth2 credentials to retrieve an OAuth2 access token. Then, the app uses this token to get authenticated by the backend. The tokens are solely used for this authentication, and they are valid for a limited period of time. The OAuth2 credentials are only used to issue access tokens. Whenever needed, the server uses the $PUID$ to generate and send to the app one or a batch of pseudorandom temporary IDs.

\subsubsection{Proximity tracing}

This section describes PEPP-PT NTK~\cite{PEPP-PT2020PEPP-PT-german}; ROBERT~\cite{PRIVATICS2020ROBERT} follows a similar approach.
For every period $t$, say, 1h, the backend generates a single secret key $BK_t$ shared with all users. The backend generates enough $BK_t$ keys to cover a larger period in the future, say 2 days. Then, for each user, the backend generates an ephemeral BLE ID ($EBID$) for every $t$, by encrypting their $PUID$ with the $BK_t$:

\[EBID_t(PUID) = AES(BK_t, PUID)\]

Each app broadcasts its current valid $EBID_t$ via BLE advertisements using the BLE privacy feature to prevent tracking of users who send out continuous BLE advertisements. Using this feature, temporary addresses instead of fixed hardware (MAC) addresses are transmitted. The app implementation must use a new temporary address with every new $EBID$, to avoid linking of these two IDs.

Each app also constantly scans for other BLE broadcasts from PEPP-PT apps and records the received $EBID$s, the current time and metadata of the BLE connection. The metadata include parameters like the Received Signal Strength Indicator (RSSI) and outgoing and incoming signal levels (TX/RX power), which can assist in calculating the distance between the two communicating smartphones. The above data are stored only on the smartphone for as long as the user is not infected, and they are deleted after the epidemiological relevant time, say, 21 days.

\subsubsection{Sharing proximity data with the server}

When a user is tested as infected, the collected data are sent to the backend for assessing which other users are at risk and notify them. The backend holds these data for up to 3 weeks. To upload the data to the backend, a healthcare professional provides a Transaction Authentication Number (TAN) to the infected user by out-of-band means. The backend associates each $EBID$ received with its corresponding $PUID$ and calculates the risk for the $PUID$ holder.

To protect the privacy of infected users from eavesdroppers, in the NTK proposed implementation, the backend pushes notifications to infected as well as a random number of other user apps. The push notification acts as a trigger for the app to send a pull request to the backend. For users at risk, the pull request returns information to the user about potential infection and instructions. For the rest of the users the exchanged messages are just ``noise'' and no information or instructions are provided by the app. In ROBERT, a pure pull approach is followed where the app regularly inquires the backend server with its $EBID$s. According to the risk assessment procedure run on the server, the app pulls a notification informing the user whether they are at risk or not.

\subsubsection{Federation with other backends}

The federation of PEPP-PT with other systems is considered out of scope of the specification; however, some general guidelines are provided. To facilitate the federation of backend services, it is only necessary for a backend to recognise the originating backend of an $EBID$. This can be achieved by including an Encrypted Country Code (ECC) into the $EBID$ so that, for example, the ECC consumes 1 byte out of the 16 bytes available for the $EBID$. When a foreign backend receives an $EBID$ that does not belong to it, it just forwards it to the home backend. The home backend is responsible to determine how the $BK_t$ keys and the $EBID$ are constructed, as well as how the risk analysis is performed.

\subsection{Other frameworks}
While the previous frameworks represent the main contact tracing approaches nowadays, additional solutions have been recently proposed, and they are described below.

\paragraph{BlueTrace~\cite{BlueTrace2020BlueTrace}}
This framework represents the approach used by the TraceTogether app \cite{TraceTogether2020TraceTogether}, which was initially developed by Singapore's Government Technology Agency and Ministry of Health. The approach follows a centralised solution in which users register their phone numbers in the backend service, which provides random IDs that are associated with such numbers. These IDs are used during smartphones' encounters. However, in case a user is infected, they will be authorised to share their encounter history with the health authority, which in turn can obtain the phone number of the infected person, and the number of people who were in contact with them. Therefore, based on the BlueTrace design, the backend service is able to access users' personally identifiable information. 

\paragraph{TraceSecure~\cite{Bell2020TraceSecure}}
In this case, two alternative solutions are proposed. The first is based on the framework provided by the TraceTogether app and employs public key cryptography. Specifically, the authors define two versions requiring two or three separate non-colluding parties who administer the system that can be associated to health authorities or other government services. The second approach is based on the use of homomorphic encryption by leveraging the ability of the parties to exchange secrets for the sake of providing advanced privacy features. 

\paragraph{DESIRE~\cite{2020DESIRE}}
This system has been recently proposed by INRIA and integrates different aspects of centralised and decentralised models. Specifically, DESIRE follows the ROBERT architecture in which risk scores and notifications are handled by a server. Nevertheless, the approach relies on the concept of Private Encounter Tokens (PETs), which are privately generated by users, and thus are unlinkable. The server is intended to match the PETs provided by infected users with the PETs of requesting people. Furthermore, the information hosted by the server is encrypted with cryptographic keys, which however are locally stored in the users' smartphones.

\paragraph{TCN~\cite{2020TCN}}
The Temporary Contact Numbers (TCN) is a decentralised contact tracing protocol based on the exchange of 128-bit temporary IDs among nearby smartphones using BLE. These IDs are pseudorandom and generated locally on the smartphone. When a user develops symptoms or is diagnosed as infected, the app can upload a report with the collected IDs to a central portal. Users' apps connect periodically to the portal to check if their ID has been reported by an infected user. One of the characteristics of TCN is that the involvement of a health authority is optional. If a health authority is involved, then the test results are verified by a signature from the health authority to guarantee the report's integrity. If not, the user creates a self-report of their symptoms to inform other users who have been in proximity.

\paragraph{PACT (UW)~\cite{AllenSchoolNews2020Privacy}, \cite{Chan2020PACT}}
The approach of the Privacy-sensitive protocols And mechanisms for mobile Contact Tracing (PACT) is mainly developed by researchers from the University of Washington and is based on the definition of a third-party free framework for mobile contact tracing. Particularly, the authors define a set of protocols to strengthen privacy aspects by keeping users' data in their smartphones. Indeed, this approach is related to the DP-3T system, as only infected people will be enabled to share their data on a voluntary basis. 

\paragraph{PACT (MIT)~\cite{PACTPACT}}
The Private Automated Contact Tracing (PACT) protocol has been mainly developed by researchers of the Massachusetts Institute of Technology (MIT) and bears similarities with other relevant decentralised solutions like TCN, DP-3T, and PACT (UW). The user's app generates locally pseudorandom IDs, called chirps, which change every few minutes. The chirps are broadcast using BLE and stored locally in the phone for up to 3 months. The receiving apps can store chirps for up to 3 months as well; optionally, the receiving smartphone can also store the location of the encounter. The upload of collected chirps from infected individuals to a central database is authorised by health professionals via one-time passwords. Regularly, the apps download the database locally and check if the chirps contained in the database are present in their local contact list as well.

\paragraph{OpenCovidTrace~\cite{2020OpenCovidTrace}}
This is an open-source platform following a decentralised approach. Its aim is to integrate the most popular contact tracing protocols based on BLE and provide additional features on top of them. Such protocols include DP-3T, Google/Apple Exposure Notification, and BlueTrace. This integration is envisaged to facilitate interoperability between open-source, say, DP-3T and proprietary platforms, including Google/Apple Exposure Notification and BlueTrace. OpenCovidTrace follows the original DP-3T specifications. When the Google/Apple framework is used, pseudorandom temporary IDs are generated locally on the user's smartphone, following the DP-3T approach. If a user reports COVID-19 symptoms, the app sends to a central server the keys used to generate the temporary IDs and the location (i.e., city) of the user in the last 14 days. Periodically, the user's app downloads the keys of users reporting symptoms from the server and information on who has been in the same area with the requesting user. If a match is found, the user is notified as potentially being at risk. The BlueTrace is not yet supported by OpenCovidTrace, but it reportedly will in its next release.
 
\paragraph{Whisper Tracing~\cite{Loiseau2020Whisper}}
This  protocol follows a decentralised approach. The temporary IDs are periodically generated in the user's smartphone and exchanged with nearby smartphones over BLE. When a user is infected, the app uploads to a central server the seeds that were used to produce the temporary IDs of the last 2 weeks. No health authority is involved in certifying the infection and no authorisation is foreseen to upload the collected temporary IDs. Each user's app is sporadically synchronised with the central server and, when there are new keys, they are downloaded and processed locally to produce the temporary IDs of infected users. If a match is found, it means that the local user has been in contact with an infected user; an algorithm to estimate the exposure risk is out of the scope of the protocol. To optimise the process of temporary ID downloading, the authors propose to include a location dimension to the uploaded data, so that a user only downloads temporary IDs from infected users that have visited the same geographic area.

\subsection{Summary of contact tracing frameworks}

One common aspect of all the reviewed frameworks is the technology used for exchanging the IDs among user smartphones, which is BLE. There are some common characteristics of the aforementioned frameworks related to whether they are centralised or decentralised, the main one being what kind of information is exchanged between the app and the backend server. Namely, in decentralised frameworks, the app of an infected person uploads to the backend its temporary IDs (or the seeds to regenerate them) and each app downloads the list of these IDs.
On the other hand, in centralised frameworks, the app of a tested positive user uploads its collected IDs to the server, and only the apps of potentially infected users receive a notification with further instructions. Some apps do receive dummy traffic as a measure against traffic analysis, but in this case no notification is shown to the user.

In Table~\ref{table:frameworks}, the main aspects of the contact tracing frameworks presented above are summarised based on five diverse criteria. The ``Approach'' column shows whether a centralised or decentralised approach is followed. The ``Source code'' column demonstrates whether an open source or proprietary implementation is already available, or no implementation is available. The ``Health authority'' column shows if such an authority has been considered in the system design and whether its presence is compulsory or optional. The ``Location data collected'' column describes if it is necessary for the correct operation of the framework the collection of location data by the smartphone. Finally, ``Self-reporting'' indicates whether it is possible for an infected user to directly inform the rest of the users without the involvement of a health authority certifying the infection or not.

It should be noted that some works have been recently proposed to analyse and compare some of these approaches. In particular, \cite{Vaudenay2020Centralized} discusses the main vulnerabilities and advantages of both centralised and decentralised solutions systematically. Furthermore, \cite{FraunhoferAISEC2020Pandemic} analyses DP-3T, PEPP-PT NTK, and ROBERT from a privacy perspective.

\begin{table}
\centering
\begin{threeparttable}
\caption{Main characteristics of contact tracing frameworks.}\label{table:frameworks}

\begin{tabular}{|l|l|l|c|c|c|}
\hline
\multicolumn{1}{|c|}{\multirow{2}{*}{\textbf{Framework}}} & \multicolumn{1}{c|}{\multirow{2}{*}{\textbf{Approach}}} & \multicolumn{1}{c|}{\multirow{2}{*}{\textbf{Source code}}} &\textbf{Health} & \textbf{Location data} & \multirow{2}{*}{\textbf{Self-reporting}} \\
& & & \textbf{authority} & \textbf{collected} &\\
\hline
\hline
DP-3T & Decentralised & Open~\cite{DP3TDP3T} & Yes & No & No\\
\hline
Google/Apple & Decentralised & Proprietary\tnote{1} & Yes & No & No\\
\hline
PEPP-PT NTK & Centralised & Open\tnote{2} & Yes & No & No\\
\hline
ROBERT & Centralised & Open~\cite{GitLabStopCovid} & Yes & No & No\\
\hline
BlueTrace & Centralised & Open~\cite{OpenTrace} & Yes & No & No\\
\hline
TraceSecure & Centralised\tnote{3} & Not available & Yes & No & No\\
\hline
DESIRE & Hybrid\tnote{4} & Not available & Yes & No & No\\
\hline
PACT (UW) & Decentralised & Open~\cite{CovidSafe} & Yes & No & No\\
\hline
PACT (MIT) & Decentralised & Not available & Yes & Optional & No\\
\hline
TCN & Decentralised & Open~\cite{TCN} & Optional & No & Yes\\
\hline
Open Covid Trace & Decentralised & Open~\cite{OpenCovidTraceSource} & Yes & Yes\tnote{5} & No\\
\hline
Whisper Tracing & Decentralised & Not available & No & Optional & Yes\\
\hline
\end{tabular}

  \begin{tablenotes}
     \item[1] Proprietary API; open source reference implementations are provided for a server~\cite{gaen-server} and an Android app~\cite{gaen-app}.
     \item[2] The goal of the consortium is to open-source the code but the repositories are not available yet~\cite{PEPP-PT2020PEPP-PT}; open-source reference implementation is provided for an Android app~\cite{ntk-core-android, ntk-sample-android}.
     \item[3] While the approach is based on TraceTogether, it adds additional mechanisms (e.g., homomorphic encryption) to improve privacy aspects. 
     \item[4] It integrates some of the advantages of centralised and decentralised approaches. While users do not need to register as in the case of BlueTrace, the backend server still is able to match infected and requesting users based on PET and the risk score computation.
     \item[5] When the Google/Apple protocol is used.
   \end{tablenotes}
   
\end{threeparttable}
\end{table}
\section{Adversarial model}\label{sec:threat}
This section details on the adversaries (Adv) and their capabilities
and discusses the most relevant attacks. It also identifies the assets and any assumption made. The analysis is mostly done in a generic way, that is, it is not focused on a specific digital contact tracing architecture, namely centralised~\cite{PEPP-PTPan-European,PRIVATICS2020ROBERT}, semi-centralised~\cite{2020DESIRE}, or decentralised~\cite{Apple2020Privacy-Preserving,Troncoso2020Decentralized}. To this matter, the interested reader can refer to the heretofore relevant work examining the pros and cons of each approach~\cite{Vaudenay2020Centralized,PRIVATICS2020Proximity,TheDP3TConsortium2020DESIRE,TheDP3TProject2020SecurityPEPP,TheDP-3TProject2020SecurityROBERT,Vaudenay2020Analysis,DP-3TProject2020Response,Avitabile2020Towards}. Nevertheless, given that from a practical standpoint, i.e., in regard to the available API/SDK, the most pertinent solution for implementing such a service is the Google/Apple framework, the current section assumes that the advertisement service (beaconing\footnote{The terms ``advertisement'' and ``beacon'' are used interchangeably in this section.}) is based on BLE, in particular the ``Exposure Notification'' framework~\cite{DP3TSDKandroid,DP3TSDKios,Apple2020Exposure}.

\subsection{Adversaries}

Adversaries are individuals, groups, or organisations who attempt to compromise the security/privacy of the contact tracing service or disrupt its operation. The model considers a (i) passive or active, (ii) honest-but-curious or malicious, and (iii) outsider or insider Adv who might try to put in place the following line of actions in order to attack a digital contract tracing system:

\begin{enumerate}
    \item intercept, block, modify, inject, or replay any message in the public communication channel;
    \item use the mobile app to access the system and enable or disable the app’s notification service at will;
    \item where applicable, the same Adv can try to register with the service multiple times, i.e., create multiple profiles;
    \item the same Adv can carry multiple devices (smartphones) and install the app(s) on each of them;
    \item the same Adv can try to install the legitimate app along with custom-made ones on the same device;
    \item install high-power antennas to amplify their reception and transmission signal to cover a wider area with the purpose of magnifying their tracking or broadcasting capacity;
    \item access the data stored in the device;
    \item cause Denial of Service (DoS), commotion to the system, harassment to the end-user, or contaminate the data provided to the system;
    \item setup and operate their own rogue backend server;
    \item have access to the source code of the backend server;
    \item collude with other end-user(s), persons working for the health or law enforcement authorities, or backend server admins;
    \item compromise a backend server and the underlying IT infrastructure;
    \item trick/lure end-users into installing malware on their devices, say, by exercising social engineering techniques.
\end{enumerate}

\subsection{Assumptions}

The following assumptions are made:
\begin{itemize}
    \item The Adv adheres to all cryptographic assumptions, e.g., they are unable to decrypt a properly encrypted message without knowing the decryption key.
    \item All the communication channels between the backend server and any health authority and between the server and the end-users are secured, say, by means of a TLS tunnel and the use of strong ciphersuites. This also means that the server holds and is associated with the expected valid X.509 certificate or public key. Assuming the use of Domain Name System Security Extensions (DNSSEC), an alternative method is for clients to obtain authenticated data directly from zone operators, say, by means of the DNS-based Authentication of Named Entities (DANE) protocol~\cite{Schlyter2012DNS-Based}.
    \item For interoperable contact tracing systems, say, between different countries, it is assumed that the respective inter-communication links are secured either at the transport layer (TLS) or preferably the network layer (IPsec).
    \item Network perimeter security is enabled on the system’s IT infrastructure, including the backend server and its subsystems, and the respective systems of the involved authorities.
    \item As per the Kerckhoffs's desideratum, and as a rule of thumb, the implementers must avoid a security by obscurity strategy. Therefore, it is assumed that the source code of the app along with the technical specifications of the system are publicly available. The same must apply to any server-side code.
    \item The client-side app only requests the minimum set of permissions necessary for its operation.
    \item Participation is fully voluntary (opt-in). The user can at any time uninstall the app, deny providing its observed interactions, etc.
\end{itemize}

\subsection{Types of adversaries}
We consider the following categories of adversaries:

\begin{itemize}
    \item Non-tech-savvy: They have access to the app and may be interested in pieces of data about other users that possibly leak from the app via the user interface, the app’s internal storage, or otherwise. Such opponents are basically semi-honest, also known as honest-but-curious\footnote{They behave according to the protocol but are interested in learning as much information as possible.}.
    \item Advanced: They have the knowledge, technical skills, and considerable resources to exercise any attack against the system. Their goals include: DoS or harassment, identify infected persons with whom they have been in close proximity, monitor all kinds of traffic, including BLE beacons, app-backend communication, and exit nodes of mix networks, tracking users in short, mid or long run either by eavesdropping on weak constructed ephemeral IDs\footnote{For instance, the 16-byte ``Rolling Proximity Identifier'' contained in the ``Exposure Notification Service'' payload as given in the respective Google/Apple framework \cite{Apple2020Privacy-Preserving}.} or pseudonyms or other permanent IDs like the device’s MAC address, compromise the backend server, etc. This category embraces any kind of hacker, researcher, IT security professional, etc. These actors are supposed to be mostly malicious, but in certain cases, they can also be honest-but-curious, e.g., academic researchers.
    \item Powerful: They comprise advanced adversaries with unlimited resources, hence they are in position to exercise any kind of attack, including persistent ones, in large-scale, say, conduct tactical espionage operations or infiltrate and obtain complete control over the system’s IT infrastructure. They typically seek to learn information and extrapolate conclusions about the population of users, track specific targets (individuals) of interest or even construct the social interaction graph of a large part of the population, sabotage, cripple or paralyse the system, broadcast a surge of bogus notifications that would cause panic, undermine the credibility of the system and subvert the authorities, etc. This category encompasses state-sponsored actors and large organisations.
    \item Peripheral: If motivated properly, say, monetary gain, bribe, revenge, corruption, extortion, hacktivism, etc., these insiders might act individually or, more likely, collude with the previous two categories of adversaries (outsiders) to inflict damage or exfiltrate confidential information. This category includes members of the family, system administrators, and persons working for the health, law enforcement, or other authorities, and thus their capacity depends on their role in the organisation. For example, they may have admin access to a centralised architecture, be able to obtain subpoenas, counterfeit diagnosis tests, and so forth, granting them privileged access and elevated power. Such actors are mainly classified as honest-but-curious with more legitimate information available, but malicious behavior is not to be ruled out, e.g., think of a dissatisfied employee, a paid-off official, etc.
\end{itemize}

\subsection{Data and other assets}
The following key assets are identified:

\begin{itemize}

    \item Any piece of data stored on or leaked from the smartphone, the protocol, say, BLE, or the app. These include the received and transmitted ephemeral IDs or pseudonyms, timestamp of interactions, device or service or app IDs like MAC address, IP address, and BLE exposure notification service metadata, say, the transmission power used to calculate the distance between the devices.
    \item Any piece of data stored on or leaked from any server in the system.
    \item Any information that can be inferred by eavesdropping on wireless or wired communication links.
    \item The relevant IT infrastructure, including the machines along with any network asset.
    
\end{itemize}

\subsection{Specific attacks}

Attacks that directly stem from the voluntary use of the app, e.g., disable notifications, or others that their impact is rather minor, e.g., inferring individuals who have installed and use the app, are out of scope of this section. The interested reader may also refer to the relevant work conducted by the ROBERT~\cite{PRIVATICS2020ROBERT}, DESIRE~\cite{2020DESIRE}, DP-3T~\cite{Troncoso2020Decentralized} and other teams~\cite{Vaudenay2020Analysis,Avitabile2020Towards,Pietrzak2020Delayed,Kolias2017Breaking}. The potential attack scenarios
described below are related
to all frameworks except when explicitly mentioned otherwise, e.g., some attacks apply to centralised frameworks only.


\begin{enumerate}[label=A\arabic*.]

\item \emph{Identify infected persons} (with whom the Adv was in proximity): The Adv, being also user of the legitimate app, needs to keep an external log of their personal interactions over time, i.e., who they met. If a notification about an infected person arrives, say from the backend server, then the Adv tries to match its log, say, notes or pictures, against the data in the notification, i.e., the ephemeral IDs of the infected person and the corresponding timeframe. To this end, the Adv may register and use multiple accounts (sybils) in the system and rotate frequently between them. Thereby, they can narrow down the list of possibly infected individuals or even directly identify the infected person. Micropayments, captchas, and similar methods can alleviate the problem of creating multiple accounts, but all these antidotes can be easily outsmarted by a motivated Adv.
\item \emph{Identify areas that infected persons frequently move or live}: The Adv uses a long-range antenna connected to their device and wanders around certain areas of interest at specific times of the day, say, at night, to collect the respective ephemeral IDs. Then, they combine the notifications received against the collected data on its app to geographically map the infected individuals. In an alternative scenario, the Adv installs passive BLE receivers (readers) in several different strategic geographical areas to collect – and possibly upload to a server - as many ephemeral IDs along with the corresponding timestamps as possible. Then, they try to correlate the collected data against those included in the received notifications for infected persons. Such a strategy is effective at least for the relevant app’s contagion time window, say, 14 days. The magnitude of this strategy is augmented if the Adv colludes with a peripheral actor, e.g., a backend server administrator. A proof-of-concept implementation of such a BLE contact tracing sniffer is given in~\cite{Seiskari2020BLE}.
\item \emph{Injection of false information}: The Adv attaches a long-range antenna to their transmitting device and places it in crowded areas. In this way, they can transmit their own ephemeral IDs to a much longer distance than that of the typical transmission range of the BLE advertisement. Therefore, many more devices will perceive and log the Adv's ephemeral ID(s). Next, the Adv must flag its status as ``infected''. This can be achieved in different ways, including going to the hospital, which verifies their infection (if already), bribes an infected individual to bring the Adv's smartphone to the hospital, colludes with the health authorities, compromises the backend server, etc. A different flavor of the same attack may arise if the Adv achieves to directly compromise the backend server or collude with peripheral actors, who will enable the Adv to directly transmit bogus notifications or events. Another possibility is to turn a short encounter with a user (that would not be relevant for a COVID-19 infection) into a longer encounter that might be deemed as relevant by the risk scoring algorithm.
\item \emph{Beacon proxying}: The Adv simply relays interactions (beacons) gathered with smartphones whose end-users have high probability of being diagnosed as infected. The Adv may for example capture and immediately or later relay elsewhere all interactions captured from individuals entering or exiting a hospital or other medical facility offering COVID-19 testing. In a similar scenario, the Adv collects a plethora of ephemeral IDs and uses a long-range antenna to replay (broadcast) it to crowded places. In this way, the receiving apps will store and possibly report in case of an infection falsified data, causing at the very least commotion to the system, undermining its credibility.
\item \emph{Beacon wormholing}: The Adv uses a custom-made app that collects beacons in a certain location. At the same time or later, they send the collected beacons over the Internet to another device for re-transmitting them in a different area. In such a scenario, the system is forced into believing that a given individual(s) was in proximity with certain persons in a disparate geographical area. By combining this strategy with that in A4 and exercising it in a broad scale would subvert the trustworthiness of the system. 
\item \emph{Tracking of individuals}: The Adv may attempt to track certain users within range via permanent IDs possibly leaked from the device, the underlying service, or the app. Such IDs include the MAC or IP address, cellular IDs like IMSI, TMSI, GUTI, etc. In particular, BLE is prone to such an attack if either the underlying operating system does not implement a robust MAC address randomisation scheme (in this case of type ``random non-resolvable'') or the synchronisation between MAC address randomisation and Bluetooth IDs is not in place\footnote{The ephemeral ID is periodically renewed to prevent location tracking of users and is broadcast via BLE advertisements using the BLE privacy feature. This feature is available as of Bluetooth 4.0 and uses regularly changing private addresses instead of fixed hardware addresses to prevent tracking of users who send out continuous BLE advertisements. The implementation must ensure that whenever a new ephemeral ID is used, the Resolvable Private Address (RPK) is changed as well to avoid linking of these two IDs. The current draft specification of the ``Privacy-Preserving Contact Tracing'' (v.1.2) developed in the context of Google/Apple joint effort, defines that ``... the advertiser address rotation period shall be a random value that is greater than 10 min and less than 20 min''. Also, the same specification designates that ``The advertiser address, Rolling Proximity Identifier, and Associated Encrypted Metadata shall be changed synchronously so that they cannot be linked''.}. In the latter case, the Adv may be able to associate a new MAC with an old ephemeral ID or new ephemeral ID with old MAC. Recent work~\cite{Becker2019Tracking} has demonstrated that several state-of-the-art devices which do implement MAC randomisation as an anonymisation measure are indeed susceptible to passive tracking.
\item \emph{End-user identification}: In cases the smartphone of an end-user directly uploads proximity tracing data to, say, a backend server, the admin(s) of that server, any passive eavesdropper exercising traffic analysis on any hop of the connection, or other actor, including the Internet Service Provider (ISP), Wi-Fi or cellular provider, are able to perceive the ID and relative location of this user by means of the associated network IDs, e.g., the IP address. Note that Transport Layer Security (TLS) does not protect against traffic analysis. This threat is more significant for infected end-users, and depending on the case, can be partially or fully mitigated using IPsec tunnels, injection of dummy traffic, trusted proxies, or anonymity networks like Tor or I2P, but, say, ``Torification'' of the app is required. Note however that all such remedies typically come at a substantial cost, namely they induce a considerable overhead in communication and processing time, drain the battery faster, and increase the volume of the data transferred, which may lead to extra charges if the user employs a cellular connection. As a side note, the number of observed ephemeral IDs may be quite large, especially in periods where the lockdown measures are eased.
\item \emph{Leakage of information stored by the app}: This requires the Adv to obtain direct access to the device, e.g., members of the family, authorities, blackmailers, thieves, etc. If so, and given that apps keep locally their own broadcast IDs along with the observed ones, the Adv is in position to learn the targeted-person’s interactions and possibly track them in a certain timeframe. Precisely, the Adv can learn the corresponding ephemeral IDs, both received and broadcast, enabling them to extract useful information about the social interactions of the user, track them (via the latter IDs), or possibly, if they have the technical skills, to forge the risk score calculated by the app. This means that apps must diminish the data stored on the device to the bare minimum, and only for as long as is required. This threat can be mitigated if the relevant pieces of data are encrypted.
\item \emph{Linkability}\footnote{According to RFC 6973, unlinkability is defined as ``Within a particular set of information, the inability of an observer or attacker to distinguish whether two items of interest are related or not (with a high enough degree of probability to be useful to the observer or attacker)'' \cite{rfc6973}.}: The Adv is in position to link broadcast IDs belonging to the same infected individual. As per A6, if the smartphone of an infected user directly transmits data to the backend server, the latter is enabled to (a) learn the number of ephemeral IDs the infected person gathered during the contagious period, (b) indirectly deduce if some users were co-located, e.g., in case the same ephemeral ID is reported by two or more infected persons during the same timeframe, and (c) possible associate all different ephemeral IDs to the same persistent network ID. Third users have also the same capacity (c), given that the IDs of the infected individuals are shared. This exposure may be for the whole contagious timeframe or a part of it, say, one day, depending on the information shared to users for reconstructing an infected individual’s ephemeral IDs. For example, the Adv might observe that they came across the same infected person at 11:00 AM and 15:00 PM. This also means that with reference to A1 and assuming that the timestamps associated with the ephemeral IDs are not obfuscated, the identification of the infected person is immediate, lifting the need for the Adv to create multiple IDs in the system.
\item \emph{Use of pseudonyms}: Typically, centralised systems rely on some type of pseudonymity, namely the central server must be able to de-obfuscate ephemeral IDs to the corresponding permanent or long-term ID of the user in order to notify the corresponding at-risk device owner. That is, in such designs, a permanent ID is assigned to the end-user during the registration phase, and the backend server generates the ephemeral IDs and pushes them to the client's device. This also means that server admins, peripheral actors, and any Adv who colludes with the former entities may be in position to learn (a) which people are at-risk, and (b) deanomymise and persistently track specific users of interest in the long run. Additionally, as more and more infected individuals upload their contact history, the central server can gradually expose the social interaction graph of a considerable part of the population for a certain epoch, including non-infected users that came in proximity with at least one infected. The wealth of privacy-sensitive information~\cite{Anonymity2014} stored in such a system and the potential for function creep\footnote{Collins dictionary defines function creep as ``The gradual widening of the use of a technology or system beyond the purpose for which it was originally intended, especially when this leads to potential invasion of privacy''.}, makes it a far more alluring target for the motivated Adv, especially for powerful ones, and thus more prone to data breaches and leaks.
\item \emph{Radio jamming}: Using a radio jammer the Adv blocks device proximity interactions in a certain area. The stronger the jammer the larger the affected area.
\item \emph{Blocking}: In some centralised approaches, the ephemeral IDs are created on the server and sent to the client. Typically, a batch of future IDs is created, e.g., enough for two days. An Adv could mount a DoS attack to prevent the IDs from reaching the client. The Adv may also block the upload of the list of ephemeral IDs to the backend.
\item \emph{Resource exhaustion}: The Adv creates an upsurge of proximity events with the aim of exhausting the recipients’ smartphone computing resources, including battery, memory, and CPU. On top of that, legitimate events may be missed or dropped by the overstressed device or app. The use of a long-range antenna is anticipated to maximise the magnitude of this tactic.
\item \emph{Beacon silencing}: The Adv tries to fool a reader into thinking that a BLE beacon is remote, while it actually exists in its vicinity. This may be feasible if the transmitted power field (txPower) also known as measured power of the beacon frame is exposed. txPower is typically a factory-calibrated constant, which denotes what is the expected Received Signal Strength Indicator (RSSI) at a distance of one meter to the beacon. As already pointed out, the txPower is used along with the RSSI in the proximity calculation. For instance, if the smartphone realises that its RSSI is identical to the txPower field contained in the advertisement, it assumes that it is exactly one meter away. Pertinent is also the fact that due to the high fluctuations of the RSSI value, the proximity calculation is typically averaged by multiple signals. The Adv may transmit a flood of spoofed beacons with a greater txPower value, thus biasing the estimation of proximity toward the fake readings, which in turn yields to a faulty proximity estimation. Lately, the specification of the Google/Apple joint framework mandates the encryption of the ``Associated Encrypted Metadata'' field, which contains the txPower value. Therefore, this field can only be decapsulated after a user is diagnosed as infected and agrees to share their daily key\footnote{Called ``Temporary Exposure Key'' in the Google/Apple framework.}. Note that signal attenuation (txPower - RSSI) is one of the risk parameters for calculating the overall exposure risk~\cite{Apple2020ENExposureConfiguration}. Given the above analysis, if an amplifying antenna is being employed, any smartphone close to it, but not the ones away from it, will observe unusual strong RSSIs. Therefore, as an defensive measure, the receiving app can be instructed to at least drop ``too loud'' advertisements. Moreover, similar to typical beacons, contact tracing apps will set txPower to a constant value, meaning that at a minimum any instance of the same app will be aware of the proper txPower value. This also means that certain thresholds regarding txPower and RSSI values need to be determined.
\item \emph{Sybils}: The Adv may install more than one contact tracing apps on the same device, i.e., the legitimate one and one or more custom-made. From a receiving (reader) device’s viewpoint, these, say, two apps running on the same smartphone will be perceived as two different devices (persons). Given MAC randomisation, it is not trivial to associate the beacons stemming from these apps with the same device. In a similar approach, the Adv carries with them more than one device with one or more apps running on each device. Such a tactic, especially if combined with A3, is certain to pollute the data received by readers, create commotion to the system, and possibly taint epidemiological analysis.
\item \emph{Malware}: The Adv may lure the user into installing a spyware-like app on their device with the purpose of realising A8. Specifically, such an app will secretly monitor for, say, BLE beacons and will report any incident of sensing these beacons to a remote service controlled by the attacker. Such an app may also ask for location permissions, e.g., for activating GPS, thus enabling the Adv to profile the user in the long-term. In another scenario, the Adv may lure the victim into downloading a repackaged app instead of the legitimate one, say, by means of malvertising. Such an app may track relevant data and transmit it to the Adv via a covert channel, display fake notifications to the user, block the receiving/uploading of certain messages, forge messages and risk-scores, to name a few.
\item \emph{Man-in-the-middle}: Some systems, especially the centralised ones, require user registration prior to allowing access to the service. In such a case, the Adv, residing in the same network as the victim(s) or colluding with one who does, may attempt to gather user credentials, namely username and password by exercising a Man-In-The-Middle (MITM) attack. To this end, the Adv uses, say, the SET tool~\cite{Social-Engineer} and clones the legitimate website enabling registration at the backend on the local host running Apache server. Next, the Adv needs to redirect the victim(s) on the local network from the legitimate website to the cloned one. For this, they typically need to create a DNS file that will enable the redirection. Also, the Adv must find a way to overcome TLS protection (HTTPS) on the registration page, as well as the protection provided by the HTTP Strict Transport Security (HSTS) header, if any, which compels web browsers to enforce HTTPS. This may be achieved by using a MITM framework like Bettercap~\cite{bettercap}, which uses a built-in SSL-Strip function. It is stressed that such an attack uses publicly available tools, and thus can be mounted by even a script-kiddie.

\end{enumerate}

\subsection{Mitigation techniques}
This section summarises the main approaches to alleviate the potential attack scenarios presented above. The focus of this section is on the main centralised and decentralised frameworks, that is, DP-3T and PEPP-PT; however, the methods presented here can be applied to other frameworks with the same architecture as well.

\paragraph{DP-3T}
In regards to security, the DP-3T framework describes three main aspects: fake contact events exploited by a potential attacker to trick the user; suppressing at-risk contacts, in which people are blocked from knowing they are at risk; and prevent contact discovery, in which the system functionality is obstructed due to, say, jamming of Bluetooth radio. As discussed by the DP-3T consortium \cite{Troncoso2020Decentralized}, to cope with fake contact events, the low-cost design prevents relay attacks in which an EphID is relayed with a delay of more than one day, because the seeds of infected users are bound to the day on which they are valid. In the case of the unlinkable design, this aspect is further mitigated, because EphIDs are linked to a certain \textit{epoch}, so that the attacker should rebroadcast the EphIDs in the same epoch. 
To avoid a user claiming another user's EphID as their own, the use of a hash function and a pseudo-random function to derive EphIDs from a seed makes infeasible to learn another user's seed from observing their broadcasts.

With respect to privacy concerns, the DP-3T specification considers several aspects. First, the \textit{social graph} represents the social relationships between users in the system. The DP-3T approach does not reveal such a graph to any party, except for the two users involved in a contact. Second, the \textit{interaction graph} reflects close-range physical interactions between users. In this case, it is not possible to infer about people in contact from the EphIDs being shared. Third, \textit{location traceability} should be also avoided. In DP-3T, the EphIDs are unlinkable, and only the user's smartphone knows the seed to generate them. In case a user is infected and gives permission, the seed of the first contagious day is uploaded to the backend. Taking into account this seed, the user's EphIDs are linkable from the start of the contagious day until the seed is uploaded, when the phone will generate a new seed. In the case of the unlinkable design, EphIDs remain unlinkable, as long as the server is considered honest. Fourth, \textit{at-risk individuals} make reference to people who recently contacted with infected individuals, and only they should know about this circumstance. The system does provide this feature since the seeds of an infected person do not reveal anything about their contacts. Fifth, \textit{COVID-19 positive status} means that the system should ensure that only infected people and the corresponding health authority know about this circumstance. In the case of the unlinkable design, this issue is mitigated because of the unlinkability properties of the EphIDs of infected people.

\paragraph{PEPP-PT} 
According to its designers, malicious backend admins are not considered as adversaries because the cost to succeed in the attack outweighs the benefits.
Also, state-level adversaries are considered out of scope of the threat model of the system; a potential mitigation is that users can change their pseudonym at any time by re-installing the app and, thus, evade a continuous tracking by a state-level adversary.

Regarding Sybil attacks, i.e., registration of multiple accounts by the same user, PoW and Captcha are used. The authentication of the EBIDs is addressed by using authenticated channels between the app and the backend. By using a TAN provided by a healthcare professional it is ensured that only officially diagnosed users can upload EBID lists to the backend server. The backend server is the only one having in possession - ideally stored in a hardware security module - the secret key to produce EBIDs from the persistent ID (PUID) of the user. Thus, EBIDs are linkable to a PUID by the backend server only.
In scenario A3, the possibility of turning a short encounter with a user (that would not be relevant for a Covid-19 infection) into a longer encounter that might be deemed as relevant by the risk scoring algorithm is described. A potential solution could be to use signed EBIDs, but this would create key management issues considering the large scale of proximity tracing apps. 

Regarding the privacy properties of the system, the temporary IDs exchanged through BLE are pseudorandom and changed regularly, making it difficult for an attacker to associate multiple temporary IDs to the same device and consequently identify its user. Also, these IDs remain stored in the collecting devices and sent to the server only if the user is positive to COVID-19. The network traffic of all users when requesting updates of their risk score is indifferent, so that an eavesdropper cannot distinguish at-risk from not-at-risk users. Periodically, older data are erased, reducing the probability of data misuse. The location privacy of users is protected by not collecting location data.
An adversary can determine that a user is positive to COVID-19 by observing network traffic, namely the user uploads a higher volume of data than usual; a mitigation measure is to use mix networks like Tor.

\section{Digital contact tracing mobile apps} \label{sec:apps}

To help health authorities and governments in the fight against COVID-19 pandemic at national level, many countries decided to develop and deploy mobile apps. This work focuses on European initiatives. Their goal is to detect as soon as possible new potential sources of infection, so that the COVID-19 spread could be promptly mitigated. Two types of mobile apps can be found so far, namely contact tracing apps and location sharing apps.

A \textit{contact tracing app} is based on the digital contact tracing frameworks presented in Section~\ref{sec:frameworks} relying on proximity wireless technology such as BLE. When two users are physically close, the smartphones send their identity in terms of ephemeral IDs or pseudonyms to each other. Each smartphone records all its encounters that happened within a period of time, say, the last 14 days. If a user declares a COVID-19 infection, then all their encountered users that were evaluated at risk are warned of the situation through a central server, acting as an information dispatching office, and are requested to remain in self-isolation.

A \textit{location sharing app} relies on the smartphone positioning information, i.e., via GPS tracking or cell tower mapping. For such an app, the user needs to accept that their smartphone sends on a regular basis, say, every 5 minutes its position to a central server, which can map every user for an unlimited period of time. If a user declares a COVID-19 infection, then all the users that were within a close range to the infected user during the last, say, 14 days are warned of the situation. As with contact tracing apps, all the concerned users are requested to self-isolate.

Note that location sharing apps appear to be far less privacy-preserving for the users as the latter must agree to share continuously their location with a central server. In the case of contact tracing apps, only the - sometimes anonymised - information related to the encounters are shared.

Within the European landscape, some countries are still in the process of developing apps to help the fight against COVID-19.
For instance, the U.K. government announced that the National Health Service (NHS) digital department deployed a contact tracing app, called \emph{NHS COVID-19}~\cite{NHSX}. The app uses BLE and is available for Android 6+ and iOS 13.5+. It is based on a centralised approach and its source code can be found on GitHub~\cite{NHSXGitHub}. It is currently under a testing phase on the Isle of Wight and in the London Borough of Newham~\cite{DepartmentofHealthandSocialCare2020Coronavirus}. Note that, while the NHS COVID-19 app was still in testing, a decentralized tracing app has been developed in parallel as a backup, based on the Google/Apple framework. Eventually, the U.K. government decided to switch to the decentralised approach~\cite{DepartmentofHealthandSocialCare2020AppUpdate}.
Belgium also entered quite late in the process of developing an app, as some medias announced a release for September 2020.
Lithuania is planning to buy a contact tracing app.
Other countries like Sweden have no plans to develop or adopt any mobile app regarding the COVID-19 emergency.

Tables~\ref{table:list-apps-char} and~\ref{table:list-apps-data} sum up the different mobile apps that European countries have already deployed to fight against COVID-19. Apps that are deployed outside the European continent are left for future work. A detailed discussion on each app is provided in the subsequent subsections. In particular, we focus our analysis in the main operational aspects of each app, including installation, functioning, as well as data retention and processing aspects, which are key considerations for privacy concerns. However, it should be noted that additional aspects could affect the massive deployment of contact tracing apps. Indeed, the requirements on battery usage, and the compatibility between different apps and OS versions are explicitly mentioned by {\cite{ahmed2020survey}} as additional user concerns that could play a very significant role to roll out apps, and on the decision-making process of other countries developing such pieces of software. Our work provides a complementary analysis to existing literature by providing an exhaustive review of ongoing efforts in EU countries.

\begin{table*}
\caption{List of the European deployed mobile app characteristics.}\label{table:list-apps-char}
\centering
\resizebox{\textwidth}{!}{
\begin{threeparttable}
\begin{tabular}{|l|l|l|r|l|l|p{5cm}|}
\hline
\multicolumn{1}{|c|}{\multirow{2}{*}{\textbf{Country}}} & \multicolumn{1}{c|}{\multirow{2}{*}{\textbf{App name}}} & \multicolumn{1}{c|}{\multirow{2}{*}{\textbf{Platform}}} & \multicolumn{1}{c|}{\textbf{Downloads\tnote{1}}} & \multicolumn{1}{c|}{\textbf{Architecture}} & \multicolumn{1}{c|}{\textbf{Wireless}} & \multicolumn{1}{c|}{\multirow{2}{*}{\textbf{App providers}}}\\
& & & \multicolumn{1}{c|}{\textbf{(Google Play)}} & \multicolumn{1}{c|}{\textbf{framework}} & \multicolumn{1}{c|}{\textbf{technology}} & \\
\hline
\hline
Austria & Stopp Corona & Android 6+ & 100 000+ & Decentralised & BLE & \multirow{3}{5cm}{Austrian Red Cross, UNIQA Foundation, Accenture, Microsoft, World-Direct eBusiness} \\
& \cite{StoppCorona} & iOS 13.5+ & & Google/Apple & & \\
& & & & & & \\

\hline
Bulgaria & ViruSafe & Android 5+ & 10 000+ & Centralised & GPS & ScaleFocus \\
& \cite{ViruSafe} & iOS 10+ & & Proprietary & & \\

\hline
Croatia & Stop COVID-19 & Android 6+ & 10 000+ & Decentralised & BLE & Ministry of Health  \\
& \cite{StopCOVID19} & iOS 13.5+ & & Google/Apple & & \\

\hline
Cyprus & CovTracer & Android 5+ & 1 000+ & Centralised & \multirow{4}{1.7cm}{GPS, IP addresses, cell towers, Bluetooth} & \multirow{2}{5cm}{RISE, XM.com, Prountzos \& Prountzos LLC} \\
& \cite{CovTracer} & iOS 9+ & & Proprietary & & \\
& & & & & & \\
& & & & & & \\

\hline
Czech & eRouska & Android 5+ & 100 000+ & Centralised & BLE & Ministry of Health \\
Republic & \cite{eRouska1} & iOS 11+ & & Proprietary & & \\

\hline
Denmark & Smittestop & Android 6+ & 100 000+ & Decentralised & BLE & \multirow{6}{5cm}{Danish Ministry of Health and Elderly, Danish Agency for Patient Safety, Danish Health and Medicines Authority, Statens Serum  Institut, Danish Digitization Agency, Netcompany} \\
& \cite{Smittestop} & iOS 13.5+ & & Google/Apple & & \\
& & & & & & \\
& & & & & & \\
& & & & & & \\
& & & & & & \\

\hline
Estonia & Hoia & Android 6+ & 50 000+ & Decentralised & BLE & \multirow{5}{5cm}{Estonian Health Board, Ministry of Social Affairs, Health and Welfare Information Systems Centre, voluntary consortium of Estonian companies} \\
& \cite{Hoia} & iOS 13.5+ & & Google/Apple & & \\
& & & & & & \\
& & & & & & \\
& & & & & & \\

\hline
Finland & Koronavilkku & Android 6+ & 1 000 000+ & Decentralised & BLE & \multirow{5}{5cm}{Finnish Institute for Health and Welfare, Ministry of Social Affairs and Health, Social Insurance Institution of Finland, SoteDigi Oy, Solita Oy}\\
& \cite{Koronavilkku} & iOS 13.5+ & & Google/Apple & & \\
& & & & & & \\
& & & & & & \\
& & & & & & \\

\hline
France & StopCovid & Android 5+ & 1 000 000+ & Centralised & BLE, & \multirow{3}{5cm}{INRIA, Ministry for Solidarity and Health, Ministry of State for Digital Affairs} \\
& \cite{StopCovid} & iOS 11.4+ & & ROBERT & ultrasounds & \\
& & & & & & \\

\hline
Germany & Corona-Warn-App & Android 6+ & 5 000 000+ & Decentralised & BLE & \multirow{2}{5cm}{Deutsche Telekom, SAP Deutschland} \\
& \cite{CoronaWarnApp} & iOS 13.5+ & & Google/Apple & & \\

\hline
Hungary & VirusRadar & Android 5+ & 10 000+ & Centralised & BLE & NextSense \\
& \cite{VirusRadar} & iOS 11+ & & Proprietary & & \\

\hline
Ireland & COVID Tracker & Android 6+ & 500 000+ & Decentralised & BLE & \multirow{3}{5cm}{Department of Health, Health Service Executive, NearForm, Twilio, Amazon} \\
& \cite{COVIDTracker} & iOS 13.5+ & & Google/Apple & & \\
& & & & & & \\

\hline
Italy & Immuni & Android 6+ & 1 000 000+ & Decentralised & BLE & \multirow{4}{5cm}{Ministry of Health, Ministry for Innovation Technology and Digitalisation, Bending Spoons S.p.A., Sogei S.p.A.} \\
& \cite{Immuni} & iOS 13+ & & Google/Apple & & \\
& & & & & & \\
& & & & & & \\

\hline
Latvia & Apturi Covid & Android 6+ & 100 000+ & Decentralised & BLE & Ministry of Health, SPKC \\
& \cite{Apturi} & iOS 13.5+ & & Google/Apple & & \\

\hline
Netherlands & CoronaMelder & Android 6+ & 100 000+ & Decentralised & BLE & \multirow{4}{5cm}{Minister of Health, Welfare and Sport, National Institute for Health and Environment, Municipal Health Services, CIBG, KPN}\\
& \cite{CoronaMelder} & iOS 13.5+ & & Google/Apple & & \\
& & & & & & \\
& & & & & & \\

\hline
Norway & Smittestopp & Android 5+ & 100 000+ & Centralised & Bluetooth, & \multirow{2}{5cm}{Ministry of Health, Institute of Public Health, Simula} \\
& \cite{helsenorge2020Together} & iOS 12+ & & Proprietary & GPS & \\

\hline
Poland & ProteGO & Android 6+ & 500 000+ & Decentralised & Bluetooth & \multirow{2}{5cm}{Ministry of Digital Affairs, consortium of Polish companies} \\
& \cite{ProteGO} & iOS 13.5+ & & Google/Apple & & \\

\hline
Portugal & StayAway Covid & Android 6+ & 500 000+ & Decentralised & BLE & \multirow{2}{5cm}{Ministry of Health, NESC TEC, ISPUP, Keyruptive, Ubirider} \\
& \cite{StayAwayCovid} & iOS 13.5+ & & Google/Apple & & \\

\hline
Slovakia & ZostanZdravy & Android 5+ & 10 000+ & Centralised & BLE, GPS & \multirow{2}{5cm}{ZostanZdravy and Sygic initiatives, Slovak volunteers} \\
& \cite{SlovakiaZostanZdravy} & iOS 10+ & & Proprietary & & \\

\hline
Slovenia & OstaniZdrav & Android 6+ & 50 000+ & Decentralised & BLE & \multirow{2}{5cm}{National Institute of Public Health, Ministry of Public Administration} \\
& \cite{OstaniZdrav} & iOS 13.5+ & & Google/Apple & & \\

\hline
Spain & RadarCOVID & Android 6+ & 1 000 000+ & Decentralised & BLE & \multirow{5}{5cm}{General Secretariat for Digital Administration, State Secretariat for Digitalisation and Artificial Intelligence, Ministry of Economic Affairs and Digital Transformation} \\
& \cite{RadarCOVID} & iOS 13.5+ & & Google/Apple & & \\
& & & & & & \\
& & & & & & \\
& & & & & & \\

\hline
Switzerland & SwissCovid & Android 6+ & 500 000+ & Decentralised & BLE & \multirow{4}{5cm}{Federal Office of Public Health, Federal Office of Information Technology, Systems and Telecommunication} \\
& \cite{SwissCovid} & iOS 13.5+ & & Google/Apple & & \\
& & & & & & \\
& & & & & & \\
\hline
\end{tabular}

\begin{tablenotes}
    \item[1] The number of downloads are as of September 15, 2020.
\end{tablenotes}

\end{threeparttable}
}
\end{table*}

\begin{table*}
\caption{List of the European deployed mobile app data management.}\label{table:list-apps-data}
\centering
\resizebox{\textwidth}{!}{
\begin{tabular}{|l|l|p{5cm}|p{6.6cm}|p{4.5cm}|}
\hline
\multicolumn{1}{|c|}{\multirow{2}{*}{\textbf{Country}}} & \multicolumn{1}{c|}{\multirow{2}{*}{\textbf{App name}}} &  \multicolumn{1}{c|}{\multirow{2}{*}{\textbf{Data collection (server-side)}}} & \multicolumn{1}{c|}{\multirow{1}{*}{\textbf{Data retention}}} & \multicolumn{1}{c|}{\multirow{2}{*}{\textbf{Data access (server-side)}}}\\
& & & \multicolumn{1}{c|}{\multirow{1}{*}{\textbf{(h=hour, d=day, m=month, y=year)}}} & \\
\hline
\hline
Austria & Stopp Corona & \multirow{2}{5cm}{UUIDs, smartphone numbers of infected users} & \multirow{2}{6.6cm}{- UUIDs, smartphone numbers of infected users: 30d (server)} & Austrian Red Cross\\
& \cite{StoppCorona} &  & & \\
& & & - UUIDs, encounters details: 14d (smartphone) & \\

\hline
Bulgaria & ViruSafe & \multirow{5}{5cm}{Smartphone number, personal ID number or passport number, age, gender, chronic illnesses, answers to personal status questions, location of the smartphone} & \multirow{2}{6.6cm}{- All data stored for the duration of the state emergency period} & \multirow{3}{4.5cm}{Ministry of Health, authorised governmental institutions with a digital certificate, doctors} \\
& \cite{ViruSafe} &  & & \\
& & & & \\
& & & & \\
& & & & \\

\hline
Croatia & Stop COVID-19 & \multirow{2}{5cm}{UUIDs of infected users, verification codes} & - UUIDs of infected users: unknown (server) & Ministry of Health \\
& \cite{StopCOVID19} & & - Verification codes: 14d (server) &\\
& & & - UUIDs, encounters details: 14d (smartphone) & \\

\hline
Cyprus & CovTracer &  \multirow{4}{5cm}{Geolocation data of infected users (last 2 weeks), name, address, date of birth, reason(s) of moving per occasion, phone number, email, password} & - All data stored for 1y & \multirow{4}{4.5cm}{Personal data only accessible by RISE, geolocation data shared with Cypriot epidemiologists}\\
& \cite{CovTracer} &  & & \\
& & & & \\
& & & & \\
& & & & \\

\hline
Czech & eRouska & \multirow{2}{5cm}{Smartphone numbers, encounters details of infected users} & - Smartphone numbers: 6m & \multirow{2}{4.5cm}{Ministry of Health, regional health authorities}\\
Republic & \cite{eRouska1} &  & - Encounters details of infected users: 12h & \\
& & & - Data on smartphone: 30d & \\

\hline
Denmark & Smittestop & \multirow{2}{5cm}{NemIDs and UUIDs of infected users} & - NemIDs of infected users: 24h (server) & \multirow{2}{4.5cm}{Danish Agency for Patient Safety}\\
& \cite{Smittestop} & & - UUIDs of infected users: 14d (server) & \\
& & & - UUIDs, encounters details: 14d (smartphone) & \\

\hline
Estonia & Hoia & UUIDs of infected users & - UUIDs of infected users: 14d (server) & \multirow{3}{4.5cm}{Estonian Health Board, Health and Welfare Information Systems Centre}\\
& \cite{Hoia} & & - UUIDs, encounters details: 14d (smartphone) & \\
& & & & \\

\hline
Finland & Koronavilkku &  UUIDs of infected users & \multirow{2}{6.6cm}{- UUIDs of infected users: until 31/03/2021 (server)} & \multirow{2}{4.5cm}{Social Insurance Institution of Finland}\\
& \cite{Koronavilkku} & & & \\
& & & - UUIDs, encounters details: 14d (smartphone) & \\

\hline
France & StopCovid & \multirow{2}{5cm}{UUIDs, encounters details of infected users} & - Encounters details of infected users: 15d & Outscale\\
& \cite{StopCovid} & & \multirow{2}{6.6cm}{- All other data: not more than 6m after the end of the health emergency state} & \\
& & & & \\

\hline
Germany & Corona-Warn-App & UUIDs of infected users, test results & - UUIDs of infected users: 14d (server) & \multirow{2}{4.5cm}{Deutsche Telekom, SAP Deutschland}\\
& \cite{CoronaWarnApp} & & - Test results: 21d (server) & \\
& & & - UUIDs, encounters details: 14d (smartphone) & \\

\hline
Hungary & VirusRadar & \multirow{2}{5cm}{UUIDs, smartphone numbers, encounters details of infected users} & \multirow{2}{6.6cm}{- UUIDs, smartphone numbers: as long as required (server)} & \multirow{4}{4.5cm}{National Center for Public Health, Government Informatics Development Agency} \\
& \cite{VirusRadar} & & & \\
& & & \multirow{2}{6.6cm}{- Encounters details of infected users: 30d (server)} & \\
& & & & \\
& & & - UUIDs, encounters details: 14d (smartphone) & \\

\hline
Ireland & COVID Tracker & \multirow{2}{5cm}{UUIDs of infected users, smartphone numbers (optional)} & - UUIDs of infected users: 14d (server) & \multirow{2}{4.5cm}{Health Service Executive, NearForm, Twilio}\\
& \cite{COVIDTracker} & & - UUIDs, encounters details: 14d (smartphone) & \\
& & & - Smartphone numbers: as long as needed & \\

\hline
Italy & Immuni & \multirow{2}{5cm}{UUIDs, encounters details of infected users, operational data} & - All data: until 01/12/2020 & Ministry of Health, Sogei S.p.A.\\
& \cite{Immuni} & & & \\

\hline
Latvia & Apturi Covid & \multirow{2}{5cm}{UUIDs, encounters details of infected users} & - UUIDs, encounters details: 14d (smartphone) & \multirow{3}{4.5cm}{SPKC, anonymised data accessible for epidemiological research}\\
& \cite{Apturi} & & \multirow{2}{6.6cm}{- All data stored on server for the required time needed by law} & \\
& & & & \\

\hline
Netherlands & CoronaMelder & UUIDs of infected users & - UUIDs of infected users: 14d (server) & \multirow{3}{4.5cm}{Minister of Health, Welfare and Sport, Municipal Health Services}\\
& \cite{CoronaMelder} & & - UUIDs, encounters details: 14d (smartphone) & \\
& & & & \\

\hline
Norway & Smittestopp & \multirow{4}{5cm}{UUIDs, smartphone numbers, age, GPS location, operating system, version number and phone model, encounters details} & - All personal data: until 01/12/2020 & \multirow{4}{4.5cm}{Ministry of Health, anonymised data accessible to the Institute of Public Health, authorised personnel}\\
& \cite{helsenorge2020Together} & & - GPS data and encounters details: 30d & \\
& & & & \\
& & & & \\

\hline
Poland & ProteGO & UUIDs of infected users & - UUIDs of infected users: 14d (server) & Ministry of Digital Affairs\\
& \cite{ProteGO} & & - UUIDs, encounters details: 14d (smartphone) & \\

\hline
Portugal & StayAway Covid & UUIDs of infected users & - UUIDs of infected users: 14d (server) & Ministry of Health\\
& \cite{StayAwayCovid} & & - UUIDs, encounters details: 14d (smartphone) & \\

\hline
Slovakia & ZostanZdravy & \multirow{3}{5cm}{UUIDs, smartphone numbers of infected users, encounters details of infected users} & \multirow{2}{6.6cm}{- All data stored for the duration of the state emergency period} & \multirow{2}{4.5cm}{Slovak government, health authorities}\\
& \cite{SlovakiaZostanZdravy} & & & \\
& & & - Smartphone numbers: 180d & \\
& & & - Encounters details of infected users: 21d & \\

\hline
Slovenia & OstaniZdrav & \multirow{2}{5cm}{UUIDs of infected users, Covid codes} & - UUIDs of infected users: 14d (server) & \multirow{3}{4.5cm}{National Institute of Public Health, Ministry of Public Administration} \\
& \cite{OstaniZdrav} & & - teleTan codes: 21d (server) &\\
& & & - UUIDs, encounters details: 14d (smartphone) & \\

\hline
Spain & RadarCOVID & UUIDs of infected users & - UUIDs of infected users: 14d (server) & Spanish government\\
& \cite{RadarCOVID} & & - UUIDs, encounters details: 14d (smartphone) & \\

\hline
Switzerland & SwissCovid & \multirow{2}{5cm}{UUIDs of infected users, Covid codes} & - UUIDs of infected users: 14d (server) & \multirow{3}{4.5cm}{Federal Office of Public Health, anonymised data accessible to the Federal Statistical Office} \\
& \cite{SwissCovid} & & - Covid codes: 24h (server) &\\
& & & - UUIDs, encounters details: 14d (smartphone) & \\
\hline
\end{tabular}}
\end{table*}

Several European countries were very prompt to deploy mobile apps that assist in containing as much as possible the spread of the pandemic. Many have been released since the end of the summer 2020. This section is based on the available public information of the apps at the time of its writing. Consequently, note that some technical details might be missing.

\subsection{Austria}

The Austrian Red Cross in collaboration with the UNIQA Foundation and Accenture Austria developed \emph{Stopp Corona}~\cite{StoppCorona}, a free app available for Android 6+ and iOS 13.5+. The use of the app is on a voluntary basis. It is based on an anonymous contact diary that logs the various encounters via BLE. The system architecture is based on the decentralised Google/Apple API\footnote{\emph{Stopp Corona} was initially based on the DP-3T framework but was later upgraded to conform to the Google/Apple API~\cite{AustriaSwitchGA}.}. The source code of the app can be found on GitHub~\cite{StoppCoronaGitHub}.

\paragraph{Installation}
No registration or personal information is needed to install and use the app. During the app installation, a random UUID (called temporary exposure key) is generated by the app. Then the app updates this ID every day.

\paragraph{Functioning}
When two users are physically close, their smartphones send their pseudorandom ID (derived from the current UUID and renewed at least every 30 min) to each other via BLE, and record the time of the encounter and its duration. The encounter must be at a distance of less than 1,5 meters and last for more than 15 minutes. These information related to the encounters are stored for 14 days. In case of infection, a user must provide his smartphone number. He then receives from the system a unique activation code to enter into the app, so that the app sends the user’s UUIDs of the last 14 days to the app server. The app of the other users periodically downloads the new UUIDs of infected users, and exploits them to derive the infected users' pseudorandom IDs for the recent past. If one matches the IDs stored in the smartphone's memory, the app notifies the user of the risky exposure. 

\paragraph{Data retention}
The UUIDs and smartphone numbers of infected users are stored on the app server for 30 days. All the UUIDs and details of encounters are stored for 14 days on the smartphone. 

\paragraph{Data processing}
The user's consent is required for the processing of personal data. The details regarding the privacy policy of the app and its compliance with the GDPR can be found at the webpage~\cite{StoppCoronaPrivacy}. The app is controlled by the Austrian Red Cross and technically operated by Accenture, which uses Microsoft Azure cloud service as server. In order to store and process the smartphone numbers, the system uses the Austrian hosting service World-Direct eBusiness solutions GmbH.

\subsection{Bulgaria}

After approval by the Bulgarian ministry, the company ScaleFocus developed \emph{ViruSafe}~\cite{ViruSafe}, a free app available for Android 5+ and iOS 10+. The use of the app is on a voluntary basis. Contrary to a contact tracing app, ViruSafe is based on GPS location sharing to enable institutions to act accordingly in case of an emergency. The source code of the app is available on GitHub~\cite{ViruSafeGitHub}.

\paragraph{Installation}
After downloading the app, the registration with a personal ID number is required, and a SMS validation phase is performed to link the smartphone number to the user's identity.

\paragraph{Functioning}
The app has a location tracker based on GPS coordinates, enabled voluntarily by the user, to create a heatmap with potentially infected people. The users can also share any chronic diseases they may have. Additionally, the app can notify users under quarantine when the quarantine period is over. 

\paragraph{Data retention}
The data are collected in a central registry. They include the following personal data: smartphone number, personal ID number or passport number, age, gender, chronic illnesses, answers to personal status questions, location of the smartphone. The data are reportedly stored for the duration of the emergency period as defined by the state.

\paragraph{Data processing}
The user's consent is required for the processing of personal data. All collected data are accessible by the Ministry of Health, acting as data processor, and authorised governmental institutions with a digital certificate. The app also provides physicians with access to the processed data automatically. They can decide if and when to contact the users at risk and provide medical advice.

\subsection{Croatia}

The Croatian Ministry of Health developed \emph{Stop COVID-19}~\cite{StopCOVID19}, a free app available for Android 6+ and iOS 13.5+. The use of the app is on a voluntary basis. It is based on an anonymous contact diary that logs the various encounters via BLE. The system architecture is based on the decentralised Google/Apple API. The source code of the app can be found on GitHub~\cite{StopCOVID19GitHub}.

\paragraph{Installation}
No registration or personal information is needed to install and use the app. During the app installation, a random UUID (called temporary exposure key) is generated by the app. Then the app updates this ID every day.

\paragraph{Functioning}
When two users are physically close, their smartphones send their pseudorandom ID (derived from the current UUID and renewed at least every 30 min) to each other via BLE, and record the time of the encounter and its duration. The encounter must be at a distance of less than 1,5 meters and last for more than 15 minutes. These information related to the encounters are stored for 14 days. In case of infection, a user receives from an healthcare professional a unique verification code to enter into the app, so that the app sends the user’s UUIDs of the last 14 days to the app server. The app of the other users periodically downloads the new UUIDs of infected users, and exploits them to derive the infected users' pseudorandom IDs for the recent past. If one matches the IDs stored in the smartphone's memory, the app notifies the user of the risky exposure. 

\paragraph{Data retention}
All the UUIDs and details of encounters are stored for 14 days on the smartphone. The verification codes are stored on the server for 14 days. The retention time of the UUIDs of infected users that are stored on the app server is not specified: in any case, the infected users can no longer connect to the system with these UUIDs.

\paragraph{Data processing}
The user's consent is required for the processing of personal data. The details regarding the privacy policy of the app and its compliance with the GDPR can be found at the webpage~\cite{StopCOVID19Privacy}. The app is controlled by the Ministry of Health. The servers are located in Croatia and in other European countries.

\subsection{Cyprus}

Based on the Safepaths~\cite{safepaths2020Private} MIT project, the Cypriot research centre RISE developed \emph{CovTracer}~\cite{CovTracer} with the contribution of XM.com and Prountzos \& Prountzos LLC. It is free and available for Android 5+ and iOS 9+. The use of the app is on a voluntary basis. It is not a contact tracing app, but rather a location sharing app. Note that location can be established with either GPS, IP address of Wi-Fi access points, cell towers, smartphone sensor data, or Bluetooth.

\paragraph{Installation}
Insufficient information is provided.

\paragraph{Functioning}
The app starts recording the user's location via GPS. All information remains on the smartphone. In case of infection, the user can share their geolocation data, i.e., their movements during the last two weeks with the epidemiologists. This information is a simple list of times and coordinates on a JavaScript Object Notation (JSON) file; no other identifying information is sent. This file is updated every 5 minutes on the user's smartphone. The epidemiologists check this information and may act upon, e.g., evacuate areas, perform cleaning, or inform people who were in proximity with the patient. If the user wishes, the geolocations of their movements can be uploaded to CovTracer server in an anonymised form. That is, information about the user's home and any possible identification traces are removed prior to uploading.

\paragraph{Data retention}
Along with location data, other information is collected, such as the full name, address, date of birth, and reason(s) of moving per occasion. The user's name and password provided during the registration phase may also be required. A phone number and email address may also be provided. All the data are stored on RISE servers and a cloud-based database located within the EU. Data are stored for one year unless a deletion request is received or the user consents to a longer period.

\paragraph{Data processing}
The detailed privacy policy of the app can be found in the document~\cite{CovTracerPrivacy}. The user's consent is required for the processing and sharing of personal data. By default, the personal data are only accessible by RISE, and the location data are shared with the Cypriot epidemiologists.

\subsection{Czech Republic}

Within the Czech government's ``smart quarantine'' plan, the Ministry of Health developed \emph{eRouska}~\cite{eRouska1}, a free app available for Android 5+ and iOS 11+. The use of the app is on a voluntary basis. It is based on an anonymous contact diary that logs the various encounters via BLE. The source code of the app can be found on GitHub~\cite{eRouskaGitHub}.
Note that the country is currently preparing the eRouska 2.0, which will be based on the decentralised Google/Apple API.

\paragraph{Installation}
To use the app, registration with a smartphone number is required. During app installation, a random UUID is generated and assigned to the user. Note that this ID is updated on a regular basis.

\paragraph{Functioning}
When two users are physically close, the smartphones send their ID to each other, and record via BLE the time of the encounter, its duration, and the ID of the other user. To be logged, the encounter must be at a distance of less than 2 meters and last for more than 15 minutes. Upon infection, a user sends the list of all the recorded encounters to the regional health authorities, which contact and signal the infection to the encountered users. Note that, although the regional health authorities know the ID of the infected user, they cannot reveal this information to the encountered users.

\paragraph{Data retention}
All the data are stored by the Ministry of Health. The smartphone number is reportedly stored for up to 6 months. The details of encounters is stored for 12 hours. The data on the smartphone are stored for 30 days.

\paragraph{Data processing}
The details regarding the privacy policy of the app and its compliance with the GDPR can be found in the document~\cite{eRouska2}. The app uses servers in the EU and US, only for some sub-services offered by Google. The audit of the app code can be found in the corresponding report~\cite{eRouska3}. The user's consent is required for the processing and sharing of the data, which is only accessible by the Ministry of Health and the regional health authorities.

\subsection{Denmark}

In collaboration with the Danish Agency for Patient Safety, the Danish Health and Medicines Authority, the Statens Serum Institut, the Danish Digitization Agency and Netcompany, the Danish Ministry of Health and the Elderly developed \emph{Smittestop}~\cite{Smittestop}, a free app available for Android 6+ and iOS 13.5+. The use of the app is on a voluntary basis. It is based on an anonymous contact diary that logs the various encounters via BLE. The system architecture is based on the decentralised Google/Apple API. 

\paragraph{Installation}
No registration or personal information is needed to install and use the app. During the app installation, a random UUID is generated by the app. Then the app updates this ID every 15 minutes.

\paragraph{Functioning}
When two users are physically close, their smartphones send their current UUID to each other via BLE. The app registers the encounter (i.e. its duration and the distance between the two smartphones). These information related to the encounters are stored for 2 weeks.  In case of infection, a user receives from the Danish Agency for Patient Safety an infection verification number NemID to enter into the app, so that the app sends the user’s NemID and UUIDs of the last 14 days to the app server. The app of the other users periodically downloads the new UUIDs of infected users. If one matches the IDs stored in the smartphone's memory, the app notifies the user of the risky exposure when (1) the encounter duration was more than 15 minutes, (2) the encounter distance was less that 1 meter, and (3) the encounter took place within the time period in which the infected person is expected to be contagious (i.e. between 2 days before until 8 days after the first symptoms or the day the person was tested as positive).

\paragraph{Data retention}
The NemIDs and UUIDs of infected users are stored on the app server for respectively 24 hours and 14 days. All the UUIDs and details of encounters are stored for 14 days on the smartphone. 

\paragraph{Data processing}
The user's consent is required for the processing of personal data. The details regarding the privacy policy of the app can be found in the document~\cite{SmittestopPrivacy}. The Danish Agency for Patient Safety is data responsible for the app.

\subsection{Estonia}

With the help of a voluntary consortium of Estonian companies\footnote{The Estonian consortium is composed of ASA Quality Services, Cybernetica, FOB Solutions, Fujitsu Estonia, Guardtime, Heisi IT, Icefire, Iglu, Mobi Lab, Mooncascade, and Velvet.}, the Estonian Health Board developed \emph{Hoia}~\cite{Hoia}, a free app available for Android 6+ and iOS 13.5+. The use of the app is on a voluntary basis. It is based on an anonymous contact diary that logs the various encounters via BLE. The system architecture is based on the decentralised Google/Apple API. The source code of the app can be found on GitLab~\cite{GitLabHoia}.

\paragraph{Installation}
No registration or personal information is needed to install and use the app. During the app installation, a random UUID is generated by the app. Then the app updates this ID frequently.

\paragraph{Functioning}
When two users are physically close, their smartphones send their current UUID to each other via BLE. The app assesses the risk of the encounter based on its duration and the distance between the two smartphones. These information related to the encounters are stored for 2 weeks. In case of infection, a user sends via the app its UUIDs of the last 14 days to the app server. The app of the other users periodically downloads the new UUIDs of infected users. If one matches the IDs stored in the smartphone's memory, the app notifies the user of the risky exposure.

\paragraph{Data retention}
The UUIDs of infected users are stored on the app server for 14 days. All the UUIDs and details of encounters are stored for 14 days on the smartphone. 

\paragraph{Data processing}
The details regarding the privacy policy of the app can be found in the document~\cite{HoiaPrivacy}. The app (including server) is operated in the state cloud server in Estonia managed by the Estonian Health and Welfare Information Systems Centre (TEHIK). The servers are located in Estonia.

\subsection{Finland}

The Finnish Institute for Health and Welfare (THL) developed \emph{Koronavilkku}~\cite{Koronavilkku}, a free app available for Android 6+ and iOS 13.5+. The use of the app is on a voluntary basis. It is based on an anonymous contact diary that logs the various encounters via BLE. The system architecture is based on the decentralised Google/Apple API. The source code of the app can be found on GitLab~\cite{GitHubKoronavilkku}.

\paragraph{Installation}
No registration or personal information is needed to install and use the app. During the app installation, a random UUID is generated by the app. Then the app updates this ID frequently.

\paragraph{Functioning}
When two users are physically close, their smartphones send their current UUID to each other via BLE. The app assesses the risk of the encounter based on its duration and the distance between the two smartphones. These information related to the encounters are stored for 3 weeks. In case of infection, a user receives by text message a single-use unlock code to enter into the app, so that the app sends the user's UUIDs of the last 14 days to the app server. The app of the other users periodically downloads the new UUIDs of infected users. If one matches the IDs stored in the smartphone's memory, the app notifies the user of the risky exposure.

\paragraph{Data retention}
The UUIDs of infected users are stored on the app server until 31 March 2021, according to the current legislation. All the UUIDs and details of encounters are stored for 21 days on the smartphone. 

\paragraph{Data processing}
The app has been developed by Solita Oy. The server is operated and maintained by the Social Insurance Institution of Finland (Kela).

\subsection{France}

Under the supervision of the Ministry for Solidarity and Health and the Ministry of State for Digital Affairs, INRIA researchers developed \emph{StopCovid}~\cite{StopCovid}, a free app available for Android 5+ and iOS 11.4+. The use of the app is on a voluntary basis. It is based on an anonymous contact diary that logs the various encounters via BLE. It further uses ultrasounds emitted via the smartphone's speaker and microphone to reduce the number of false positives. The system is centralised and based on ROBERT~\cite{PRIVATICS2020ROBERT}. The source code of the app can be found on GitLab ~\cite{GitLabStopCovid}. Note that the CNIL published an official opinion~\cite{CNIL202004Deliberation} in favour of the StopCovid app on April 24, 2020, and confirmed the release of the app in a decision report~\cite{CNIL202005Deliberation} in favour of the StopCovid app on May 25, 2020. At the end of May 2020, a bug bounty program is launched on the YesWeHack platform to verify the robustness of the app~\cite{StopCovidBugBounty}.

\paragraph{Installation}
No registration or personal information is needed to install and use the app. During the app installation, a random UUID is generated by the server and assigned to the user. Then the app updates the ID every 15 minutes.

\paragraph{Functioning}
When two users are physically close, the smartphones  send  their  ID  to  each  other,  and  record  via BLE the time of the encounter, its duration, and the ID of the other user. To be logged, the encounter must be at a distance of less than 1 meters and last for more than 15 minutes. Upon infection, a user receives from the health authorities a QR code that can be scanned within the app, which sends the list of all the recorded encounters to the central health authorities server. The latter signals the infection to the encountered users. Note that, although the regional health authorities know the ID of the infected user, they cannot reveal this information to the encountered users.

\paragraph{Data retention}
The central server stores the details of encounters for infected users and the UUIDs of all users. The details of encounters is stored for 15 days, either on the smartphone or on the central server. All the other data should not be stored for more than 6 months after the end of the health emergency state.

\paragraph{Data processing}
The central server is hosted by Outscale, a French cloud service provider that is part of the StopCovid project team. To date, it is the only hosting provider with the SecNumCloud qualification delivered by the French cybersecurity agency ANSSI.

\subsection{Germany}

Under the supervision of the German Federal Government and the Robert-Koch-Institut, Deutsche Telekom and SAP Deutschland developed \emph{Corona-Warning-App}~\cite{CoronaWarnApp}, a free app available for Android 6+ and iOS 13.5+. The use of the app is on a voluntary basis. It is based on an anonymous contact diary that logs the various encounters via BLE. The system architecture is based on the decentralised Google/Apple API\footnote{Germany initially backed the PEPP-PT centralised approach but later switched to the Google/Apple API~\cite{GermanySwitchDecentralised}.}. The source code of the app can be found on GitHub~\cite{CoronaWarnAppGitHub}.

\paragraph{Installation}
No registration or personal information is needed to install and use the app. During the app installation, a random UUID (called temporary exposure key) is generated by the app. Then the app updates this ID every day.

\paragraph{Functioning}
When two users are physically close, their smartphones send their pseudorandom ID (derived from the current UUID and renewed every 15 min) to each other via BLE. The app assesses the risk of the encounter based on its duration and the distance between the two smartphones. This is estimated from the signal attenuation of BLE. These information related to the encounters are stored for 2 weeks. In case of infection, a user sends via the app its UUIDs of the last 14 days to the app server. The app of the other users periodically downloads the new UUIDs of infected users, and uses them to derive the infected users' pseudorandom IDs for the recent past. If one matches the IDs stored in the smartphone's memory, the app notifies the user of the risky exposure. Optionally, if a user has been tested for the COVID-19, they can register the test in the app by scanning a QR code received from the testing laboratory. In that case, the result of the test is sent directly from the laboratory to the app server, which registers the result and sends it to the user via the app.

\paragraph{Data retention}
The UUIDs of infected users are stored on the app server for 14 days. All the UUIDs and details of encounters are stored for 14 days on the smartphone. If the option is chosen, the test results are stored on the app server for 21 days.

\paragraph{Data processing}
The details regarding the privacy policy of the app and its compliance with the GDPR can be found in the document~\cite{CoronaWarnAppPrivacy}. The app (including server) is operated and maintained by Deutsche Telekom and SAP Deutschland. The servers are located in Germany or in Europe.

\subsection{Hungary}

NextSense developed \emph{VirusRadar}~\cite{VirusRadar}, a free app available for Android 5+ and iOS 11+. Originally, NextSense developed an app for North Macedonia, and offered it for free to Hungary too. The use of the app is on a voluntary basis. It is based on an anonymous contact diary that logs the various encounters via BLE.

\paragraph{Installation}
During the app installation, a UUID is generated and assigned to the user. To use the app, the registration with a smartphone number is required, and a SMS validation phase is performed to link the smartphone number to the user's UUID. The smartphone number and UUID are stored by the Hungarian government on a secure server.

\paragraph{Functioning}
When two users are physically close, their smartphones send their ID to each other, and record via Bluetooth the time of the encounter, its duration, and the ID of the other user. The encounter must be at a distance of less than 2 meters and last for more than 20 minutes. These data are stored for 2 weeks. In case of infection, a user sends the list of all the recorded encounters to the health authorities, which contact and signal the infection to the encountered users.

\paragraph{Data retention}
The UUID and smartphone number are stored on the app server as long as required by the app, or until the withdrawal of consent from the user. The encounters details of infected users are stored on the server for 30 days. All the UUIDs and details of encounters are stored for 14 days on the smartphone.

\paragraph{Data processing}
The user's consent is required for the processing of personal data. The details regarding the privacy policy of the app can be found at the webpage~\cite{VirusRadarPrivacy}. The app is controlled by the Hungarian National Center for Public Health. The server is provided and managed by the Government Informatics Development Agency (KIFU).

\subsection{Ireland}

In conjunction with the Irish Department of Health (DoH), the Irish Health Service Executive (HSE) developed \emph{COVID Tracker}~\cite{COVIDTracker}, a free app available for Android 6+ and iOS 13.5+. The use of the app is on a voluntary basis. It is based on an anonymous contact diary that logs the various encounters via BLE. The system architecture is based on the decentralised Google/Apple API. The source code of the app can be found on GitHub~\cite{COVIDTrackerGitHub}.

\paragraph{Installation}
No registration or personal information is needed to install and use the app. Note however that the user might provide its smartphone number. During the app installation, a random UUID is generated by the app. Then the app updates this ID every 10 to 20 minutes.

\paragraph{Functioning}
When two users are physically close, their smartphones send their current UUID to each other via BLE. The app registers the encounter (i.e. its duration and the distance between the two smartphones). These information related to the encounters are stored for 2 weeks. In case of infection, a user receives from the HSE a unique code to enter into the app, so that the app sends the user’s UUIDs of the last 14 days to the app server. The app of the other users periodically downloads the new UUIDs of infected users. If one matches the IDs stored in the smartphone's memory, the app notifies the user of the risky exposure when the encounter duration was more than 15 minutes and the distance was less that 2 meters. The users in contact with an infected user can also receive a phone call from HSE if they provided their smartphone number.

\paragraph{Data retention}
The UUIDs of infected users are stored on the app server for 14 days. All the UUIDs and details of encounters are stored for 14 days on the smartphone. The smartphone numbers are stored until the app service is not needed anymore. 

\paragraph{Data processing}
The user's consent is required for the processing of personal data. The details regarding the privacy policy of the app can be found in the document~\cite{COVIDTrackerPrivacy}.  The HSE and DoH are the data controllers of the app and corresponding servers. The Irish NearForm and the American Twilio have access to the app data: NearForm is the app developer and Twilio is the company sending the text messages with the infection code. Amazon Web Services (AWS) provides the cloud server storage, the processing is performed in Ireland.

\subsection{Italy}

In collaboration with the Ministry of Health and the Ministry for Innovation Technology and Digitalisation, Bending Spoons S.p.A. developed \emph{Immuni}~\cite{Immuni}, a free app available for Android 6+ and iOS 13+. The use of the app is on a voluntary basis. It is based on an anonymous contact diary that logs the various encounters via BLE. The system is based on the decentralised Google/Apple API. The source code of the app can be found on GitHub~\cite{ImmuniGitHub}.

\paragraph{Installation}
No registration or personal information is needed to install and use the app. During the app installation, a random UUID (called temporary exposure key) is generated by the app. Then the app updates this ID every day.

\paragraph{Functioning}
When two users are physically close, their smartphones send their pseudorandom ID (derived from the current UUID and renewed every 15 min) to each other via BLE. The app assesses the risk of the encounter based on its duration and the distance between the two smartphones. This is estimated from the attenuation of BLE. These information related to the encounters are stored for 2 weeks. In case of infection, a user sends via the app its UUIDs of the last 14 days to the app server. The app of the other users periodically downloads the new UUIDs of infected users, and uses them to derive the infected users' pseudorandom IDs for the recent past. If one matches the IDs stored in the smartphone’s memory, the app notifies the user of the risky exposure. Immuni also sends to the server some analytical data. These include epidemiological (i.e. details of encounters) and operational information, and are sent for the purpose of helping the National Healthcare Service\footnote{Servizio Sanitario Nazionale} to provide effective assistance to users.

\paragraph{Data retention}
All the data stored on the smartphone or on the server will be deleted by December 31, 2020.

\paragraph{Data processing}
The app server is located in Italy and managed by Sogei S.p.A., a public Italian company. The Ministry of Health is the body that collects the data and decides for which purposes to use it. The data is used solely with the aim of containing the COVID-19 epidemic or for scientific research.

\subsection{Latvia}

In collaboration with Ministry of Health, the SPKC\footnote{The Latvian Center for Disease Prevention and Control} developed \emph{Apturi Covid}~\cite{Apturi}, a free app available for Android 6+ and iOS 13.5+. The use of the app is on a voluntary basis. It is based on an anonymous contact diary that logs the various encounters via BLE. The system is based on the decentralised Google/Apple API. The source code of the app will soon be released on GitHub~\cite{ApturiGitHub}.

\paragraph{Installation}
No registration or personal information is needed to install and use the app. During the app installation, a random UUID (called temporary exposure key) is generated by the app. Then the app updates this ID every day.

\paragraph{Functioning}
When two users are physically close, their smartphones send their pseudorandom ID (derived from the current UUID and renewed every 15 min) to each other via BLE, and record the time of the encounter and its duration. The encounter must be at a distance of less than 2 meters and last for more than 15 minutes. These information related to the encounters are stored for 14 days. In case of infection, a user sends via the app its UUIDs of the last 14 days to the app server. The app of the other users periodically downloads the new UUIDs of infected users, and exploits them to derive the infected users' pseudorandom IDs for the recent past. If one matches the IDs stored in the smartphone's memory, the app notifies the user of the risky exposure. Note that a smartphone can optionally be provided in the app, allowing the SPKC to directly contact the users in case of contact with an infected user.

\paragraph{Data retention}
The UUIDs and details of encounters are stored for 14 days on the smartphone. On the app server side, all the data are reportedly stored for the required time needed for the fulfillment of the obligations specified in regulatory enactments.

\paragraph{Data processing}
The details regarding the privacy policy of the app and its compliance with the GDPR can be found in the document~\cite{ApturiPrivacy}. The app (including server) is operated and maintained by the SPKC. The servers are located in Europe.
The following data are available to the SPKC: smartphone number of the encountered user, details of the encounter, except the UUIDs, i.e. date of contact, signal strength, duration of contact. Anonymised data can be shared for the purpose of epidemiological research.

\subsection{Netherlands}

In collaboration with the National Institute for Health and Environment (RIVM) and the Municipal Health Services (GGD), the Dutch Ministry of Health, Welfare and Sport developed \emph{CoronaMelder}~\cite{CoronaMelder}\footnote{Note that an unofficial Dutch app, called \emph{PrivateTracer}~\cite{PrivateTracer}, has been developed at the beginning of the pandemic as an initiative of the non-profit, open-source, public-private partnership PrivateTracer.org. It is based on the DP-3T approach and its source code can be found on GitLab~\cite{PrivateTracerGitLab}.}, a free app available for Android 6+ and iOS 13.5+. The use of the app is on a voluntary basis. It is based on an anonymous contact diary that logs the various encounters via BLE. The system is based on the decentralised Google/Apple API. The source code of the app can be found on GitHub~\cite{CoronaMelderGitHub}.

\paragraph{Installation}
No registration or personal information is needed to install and use the app. During the app installation, a random UUID is generated by the app. Then the app updates this ID every 15 minutes.

\paragraph{Functioning}
When two users are physically close, their smartphones send their current UUID to each other via BLE and record the time of the encounter and its duration. The encounter must be at a distance of less than 2 meters and last for more than 15 minutes. These information related to the encounters are stored for 2 weeks. In case of infection, a user receives from the GGD an alphanumeric code to enter into the app, so that the app sends the user’s UUIDs of the last 14 days to the app server. The app of the other users periodically downloads the new UUIDs of infected users. If one matches the IDs stored in the smartphone's memory, the app notifies the user of the risky exposure.

\paragraph{Data retention}
The UUIDs of infected users are stored on the app server for 14 days. All the UUIDs and details of encounters are stored for 14 days on the smartphone. 

\paragraph{Data processing}
The user's consent is required for the processing of personal data. The details regarding the privacy policy of the app can be found at the webpage~\cite{CoronaMelderPrivacy}. The app (including server) is controlled by the Minister of Health, Welfare and Sport. The server is administered by the CIBG with KPN. The user's consent is required for the processing of the data, the latter being accessible by the municipal health services (GGD).

\subsection{Norway}

The Norwegian Institute of Public Health and the Simula company developed \emph{Smittestopp}~\cite{helsenorge2020Together}, a free app available for Android 5+ and iOS 12+. The use of the app is on a voluntary basis. Smittestopp comprises an anonymous contact diary that logs the various encounters via Bluetooth and GPS location sharing. Note that Smittestopp is temporarily deactivated since June 16, 2020. Personal data stored in the central server are going to be deleted as soon as possible.

\paragraph{Installation}
Before using the app, each user must verify that their smartphone number is correctly registered in the Norwegian Contact and Reservation Register\footnote{The Norwegian Contact and Reservation Register is a national register hold by the Directorate of Digitalisation, so that the state or municipality are able to communicate directly with the Norwegian residents.}. Next, the same smartphone number will be used to communicate with the user. During the installation of the app, a random UUID is generated and assigned to the user. The smartphone number and UUID are stored on a central server.

\paragraph{Functioning}
When two users are physically close, the smartphones send their ID to each other, and record via Bluetooth the time of the encounter, its duration, and the ID of the other user. The encounter must be at a distance of less than 2 meters and last for more than 15 minutes. For more accurate positioning, the app will also record GPS coordinates. The details of the encounters logged by a smartphone along with the corresponding GPS data are sent continuously to the central server. In case of infection, a user signals it within the app, and the encountered users will receive a SMS notification of the situation. Note that the ID of the infected user is said to be kept anonymous to the encountered users.

\paragraph{Data retention}
The following data are stored: smartphone number, UUID, age, GPS coordinates, operating system, version number and phone model, details of the encounters. All the collected personal data will reportedly be stored until Dec. 1, 2020. The GPS data and any detail regarding the encounters are stored for up to 30 days in the smartphone and the server.

\paragraph{Data processing}
Data is only accessible to ``authorised personnel''. The Institute of Public Health receives anonymised data about the users’ movement patterns to monitor and analyse the effectiveness of the implemented measures against COVID-19. Independently of the app, once a user is diagnosed as infected, they will be recorded on a specific national health registry of persons tested positive to coronary infection.

\subsection{Poland}

As the result of the work of a coalition of Polish IT companies\footnote{The Polish consortium is composed of Tytani24 Sp. z o. o. (leader), The Coders Sp. z o. o., Webini Sp. z o. o., Sigma Connectivity Sp. z o. o., 25wat Sp. z o. o., Klimas Legal, Mobile Flag, and HOLDAPP.}, the Ministry of Digital Affairs developed \emph{ProteGO}~\cite{ProteGO}, a free app available for Android 6+ and iOS 13.5+. The use of the app is on a voluntary basis. It is based on an anonymous contact diary that logs the various encounters via BLE. The system architecture is based on the decentralised Google/Apple API\footnote{Initially, the app was based on the BlueTrace centralised approach; since version 4.0, it is based on the decentralised Google/Apple API~\cite{ProteGOGitHubIntro}.}. The source code of the app can be found on GitHub~\cite{ProteGOGitHub}. 

\paragraph{Installation}
No registration or personal information is needed to install and use the app. During the app installation, a random UUID (called temporary exposure key) is generated by the app. Then the app updates this ID every day.

\paragraph{Functioning}
When two users are physically close, their smartphones send their pseudorandom ID (derived from the current UUID and renewed at least every 30 min) to each other via BLE, and record the time of the encounter and its duration. The encounter must be at a distance of less than 2 meters and last for more than 15 minutes. These information related to the encounters are stored for 14 days. In case of infection, a user receives from the contact center (where the user has been positively tested) a unique PIN code to enter into the app, so that the app sends the user’s UUIDs of the last 14 days to the app server. The app of the other users periodically downloads the new UUIDs of infected users, and exploits them to derive the infected users' pseudorandom IDs for the recent past. If one matches the IDs stored in the smartphone’s memory, the app notifies the user of the risky exposure, which depends on (1) the encounter duration, (2) the encounter distance, (3) the elapsed time since the infection, and (4) the certainty of the infection.

\paragraph{Data retention}
The UUIDs of infected users are stored on the app server for 14 days. All the UUIDs and details of encounters are stored for 14 days on the smartphone. 

\paragraph{Data processing}
The details regarding the privacy and security audits of the app can be found at the webpage~\cite{ProteGOGitHubPrivacy}. The app is controlled and managed by the Ministry of Digital Affairs. The server is maintained by the National Operator Chmury Krajowej Sp. z o. o.

\subsection{Portugal}

In collaboration with INESC TEC, ISPUP, Keyruptive and Ubirider, the Portuguese Ministry of Health developed \emph{StayAway Covid}~\cite{StayAwayCovid}, a free app available for Android 6+ and iOS 13.5+. The use of the app is on a voluntary basis. It is based on an anonymous contact diary that logs the various encounters via BLE. The system architecture is based on the decentralised Google/Apple API. The source code of the app can be found on GitHub~\cite{StayAwayCovidGitHub}.

\paragraph{Installation}
No registration or personal information is needed to install and use the app. During the app installation, a random UUID (called temporary exposure key) is generated by the app. Then the app updates this ID every day.

\paragraph{Functioning}
When two users are physically close, their smartphones send their pseudorandom ID (derived from the current UUID and renewed at least every 30 min) to each other via BLE, and record the time of the encounter and its duration. The encounter must be at a distance of less than 2 meters and last for more than 15 minutes. These information related to the encounters are stored for 14 days. In case of infection, a user receives from an expert (e.g. a doctor) a unique activation code to enter into the app, so that the app sends the user’s UUIDs of the last 14 days to the app server. The app of the other users periodically downloads the new UUIDs of infected users, and exploits them to derive the infected users' pseudorandom IDs for the recent past. If one matches the IDs stored in the smartphone's memory, the app notifies the user of the risky exposure. 

\paragraph{Data retention}
The UUIDs of infected users are stored on the app server for 14 days. All the UUIDs and details of encounters are stored for 14 days on the smartphone.

\paragraph{Data processing}
The details regarding the privacy policy of the app can be found at the webpage~\cite{StayAwayCovidPrivacy}. The app is controlled by the Ministry of Health and technically operated by INESC TEC, ISPUP, Keyruptive and Ubirider. The server is hosted by the Portuguese Mint and Official Printing Office.

\subsection{Slovakia}

In synergy with the Zostanzdravy and Sygic initiatives, Slovak volunteers and experts developed \emph{ZostanZdravy} (StayHealthy)~\cite{SlovakiaZostanZdravy}, a free app available for Android 5+ and iOS 10+. The use of the app is on a voluntary basis. Like for Norway, it is a mix between anonymous contact diary that logs the various encounters via BLE advertisements and GPS location sharing.

\paragraph{Installation}
During the installation of the app, a UUID is generated and assigned to the user.

\paragraph{Functioning}
When two users are physically close, the smartphones send their ID to each other, and record via BLE beacons the time of the encounter, its duration, and the ID of the other user. For more accurate positioning, the app will also record GPS coordinates and send the corresponding anonymous logs of both users to the server. In case of infection, a user must register their smartphone number, and subsequently a SMS validation phase is performed to link the smartphone number to the user ID. The user afterwards provides the place of their quarantine so that the app can detect if the GPS location of the user is outside the quarantine area. Next, the user sends the list of all the recorded encounters to the health authorities, which communicate the infection to the encountered users.

\paragraph{Data retention}
The data related to the quarantine place and the respective GPS logs do not leave the smartphone. The general data retention is fixed for as long as required for the state emergency period. The smartphone number is stored for up to 180 days. The details of encounters is stored for 21 days.

\paragraph{Data processing}
The details regarding the privacy policy of the app and its compliance with the GDPR can be found in the document~\cite{SlovakiaZostanZdravyGDPR}. The app communicates with servers in the EU and US. The user's consent is required for the processing and sharing of the data, the latter being accessible by the Slovak government and the health authorities.

\subsection{Slovenia}

The National Institute of Public Health (NIJZ) and the Ministry of Public Administration (MJU) developed \emph{OstaniZdrav}~\cite{OstaniZdrav}, a free app available for Android 6+ and iOS 13.5+. The use of the app is on a voluntary basis. It is based on an anonymous contact diary that logs the various encounters via BLE. The system architecture is based on the decentralised Google/Apple API. The source code of the app can be found on GitHub~\cite{OstaniZdravGitHub}.

\paragraph{Installation}
No registration or personal information is needed to install and use the app. During the app installation, a random UUID (called temporary exposure key) is generated by the app. Then the app updates this ID every day.

\paragraph{Functioning}
When two users are physically close, their smartphones send their pseudorandom ID (derived from the current UUID and renewed at least every 30 min) to each other via BLE, and record the time of the encounter and its duration. The encounter must be at a distance of less than 1,5 meters and last for more than 15 minutes. These information related to the encounters are stored for 14 days. In case of infection, a user receives from the NIJZ a unique verification code (called teleTAN) to enter into the app, so that the app sends the user’s UUIDs of the last 14 days to the app server. The app of the other users periodically downloads the new UUIDs of infected users, and exploits them to derive the infected users' pseudorandom IDs for the recent past. If one matches the IDs stored in the smartphone's memory, the app notifies the user of the risky exposure. 

\paragraph{Data retention}
The UUIDs of infected users are stored on the app server for 14 days. All the UUIDs and details of encounters are stored for 14 days on the smartphone. The teleTAN verification codes are stored on the server for 21 days.

\paragraph{Data processing}
The user's consent is required for the processing of personal data. The details regarding the privacy policy of the app and its compliance with the GDPR can be found at the webpage~\cite{OstaniZdravPrivacy}. NIJZ is responsible for the app, while the app and the server are technically operated by the MJU. The server is located in Slovenia.

\subsection{Spain}

The Spanish government developed \emph{RadarCOVID}~\cite{RadarCOVID}, a free app available for Android 6+ and iOS 13.5+. The use of the app is on a voluntary basis. It is based on an anonymous contact diary that logs the various encounters via BLE. The system architecture is based on the decentralised Google/Apple API. The source code of the app can be found on GitHub~\cite{RadarCOVIDGitHub}.

\paragraph{Installation}
No registration or personal information is needed to install and use the app. During the app installation, a random UUID is generated by the app. Then the app updates this ID every 10 to 20 minutes.

\paragraph{Functioning}
When two users are physically close, their smartphones send their current UUID to each other via BLE. The app assesses the risk of the encounter based on its duration and the distance between the two smartphones. These information related to the encounters are stored for 2 weeks. In case of infection, a user receives by text message an alphanumeric code to enter into the app, so that the app sends the user’s UUIDs of the last 14 days to the app server. The app of the other users periodically downloads the new UUIDs of infected users. If one matches the IDs stored in the smartphone's memory, the app notifies the user of the risky exposure.

\paragraph{Data retention}
The UUIDs of infected users are stored on the app server for 14 days. All the UUIDs and details of encounters are stored for 14 days on the smartphone. 

\paragraph{Data processing}
The details regarding the privacy policy of the app can be found at the webpage~\cite{RadarCOVIDPrivacy}. The app is owned by General Secretariat for Digital Administration (SGAD), which is dependent of the State Secretariat for Digitalisation and Artificial Intelligence of the Ministry of Economic Affairs and Digital Transformation. The servers are located in European Union.

\subsection{Switzerland}

In collaboration with the Federal Office of Information Technology, Systems and Telecommunication (FOITT), the Swiss Federal Office of Public Health (FOPH) developed \emph{SwissCovid}~\cite{SwissCovid}, a free app available for Android 6+ and iOS 13.5+. The use of the app is on a voluntary basis. It is based on an anonymous contact diary that logs the various encounters via BLE. The system architecture is based on the decentralised Google/Apple API. The source code of the app can be found on GitHub~\cite{SwissCovidGitHub}.

\paragraph{Installation}
No registration or personal information is needed to install and use the app. During the app installation, a random UUID (called temporary exposure key) is generated by the app. Then the app updates this ID every day.

\paragraph{Functioning}
When two users are physically close, their smartphones send their pseudorandom ID (derived from the current UUID and renewed at least every 30 min) to each other via BLE, and record the time of the encounter and its duration. The encounter must be at a distance of less than 1,5 meters and last for more than 15 minutes. These information related to the encounters are stored for 14 days. In case of infection, a user receives from an expert with access rights (e.g. attending physicians) a unique activation code (called Covid code) to enter into the app, so that the app sends the user’s UUIDs of the last 14 days to the app server. The app of the other users periodically downloads the new UUIDs of infected users, and exploits them to derive the infected users' pseudorandom IDs for the recent past. If one matches the IDs stored in the smartphone's memory, the app notifies the user of the risky exposure. 

\paragraph{Data retention}
The UUIDs of infected users are stored on the app server for 14 days. All the UUIDs and details of encounters are stored for 14 days on the smartphone. The Covid codes are stored in the code management system for 24 hours.

\paragraph{Data processing}
The user's consent is required for the processing of personal data. The details regarding the privacy policy of the app can be found at the webpage~\cite{SwissCovidPrivacy}. The app is controlled by the FOPH and technically operated by the FOITT.

\section{Conclusions}
\label{sec:conclusions}

The work at hand provides a state-of-the-art and, to the best of our knowledge the first of its kind, review of the digital contact tracing apps ecosystem. Its contribution is threefold. First, it offers a succinct, but full-fledged review and classification of the hitherto complete frameworks proposed to realise such a service. Second, it details on and categorises the contact tracing apps already deployed by European countries. Lastly, it offers a generic adversary model, which not only conflates the relevant literature, but also delivers fresh perspectives to analysing such systems both from a security and data protection viewpoints. The current work can be used as a reference to anyone interested in better grasping the diverse facets of this rapidly evolving and timely area. It is also anticipated to stimulate and foster research efforts to the development of solutions that equally focus on the technological and data protection aspects.
\section{Conflict of interest}
The authors declare that there is no conflict of interest regarding the publication of this paper.

\section*{Acknowledgment}
Authors would like to thank Massimiliano Gusmini for the figures.

\bibliographystyle{IEEEtran}
\bibliography{main}

\end{document}